\documentclass[useAMS,usenatbib]{mn2e}
 \usepackage{lscape}
\usepackage{natbib}
\usepackage{epsfig}
\usepackage{rotating}
\usepackage{amsmath}
\usepackage{tabu}
\usepackage{adjustbox}
\usepackage{graphicx}

\def\lesssim{\,\lower2truept\hbox{${<\atop\hbox{\raise4truept\hbox{$\sim$}}}$}\,}
\def\gtrsim{\,\lower2truept\hbox{${>\atop\hbox{\raise4truept\hbox{$\sim$}}}$}\,}

\title[Characterization of a complete sample in the radio bands]{Characterization of polarimetric and total intensity behaviour of a complete sample of PACO radio sources in the radio bands.}
\author[V. Galluzzi et al.]{\parbox[t]{\textwidth}{
V.~Galluzzi$^{1,2}$\thanks{E-mail: vincenzo.galluzzi@unibo.it (VG)}, M.~Massardi$^{1}$\thanks{E-mail: massardi@ira.inaf.it (MM)}, A.~Bonaldi$^{3}$, V.~Casasola$^{4}$, L.~Gregorini$^{1}$, T.~Trombetti$^{1,5,6}$, C.~Burigana$^{1,5,6}$, M.~Bonato$^{1}$, G.~De Zotti$^{7}$, R.~Ricci$^{1}$, J.~Stevens$^{8}$, R.~D.~Ekers$^{8,9}$, L.~Bonavera$^{10}$, S.~di Serego Alighieri$^{4}$, E.~Liuzzo$^{1}$, M.~L\'opez-Caniego$^{12}$,
R.~Paladino$^{1}$, L.~Toffolatti$^{10,11}$, M.~Tucci$^{13}$ and J.~R.~Callingham$^{14}$.
}
\vspace*{8pt}\\
$^{1}$INAF, Istituto di Radioastronomia, Via Piero Gobetti 101, I-40129 Bologna, Italy\\
$^{2}$Dipartimento di Fisica e Astronomia, Universit\`a di Bologna, via Ranzani 1, I-40126 Bologna, Italy\\
$^{3}$Jodrell Bank Centre for Astrophysics School of Physics \& Astronomy, The University of Manchester, Manchester M13 9PL, UK\\
$^{4}$INAF - Osservatorio Astrofisico di Arcetri, Largo Enrico Fermi 5, I-50125 Firenze, Italy\\
$^{5}$Dipartimento di Fisica e Scienze della Terra, Universit\`a degli Studi di Ferrara, Via Giuseppe Saragat 1, I-44100 Ferrara, Italy\\
$^{6}$INFN-Sezione di Bologna, Via Irnerio 46, I-40126 Bologna, Italy\\
$^{7}$INAF, Osservatorio Astronomico di Padova, Vicolo dell'Osservatorio 5, I-35122 Padova, Italy\\
$^{8}$CSIRO Astronomy and Space Science, PO Box 76, Epping, NSW 1710, Australia\\
$^{9}$International Centre for Radio Astronomy Research, Curtin University, Bentley, WA 6102, Australia\\\
$^{10}$Departamento de F\'\i{sica} Universidad de Oviedo c. Federico Garc\'ia Lorca, 18, 33007 - Oviedo, Spain\\
$^{11}$INAF-IASF Bologna, Via Piero Gobetti 101, I-40129 Bologna, Italy\\
$^{12}$European Space Agency, ESAC, Camino bajo del Castillo, s/n, Urbanizaci\'{o}n Villafranca del Castillo,\\
\;\;\,Villanueva de la Ca\~{n}ada, Madrid, Spain\\
$^{13}$D\'epartement de Physique Th\'eorique and Center for Astroparticle Physics (CAP), University of Geneva,\\\;\;\, 24 quai Ernest Ansermet, CH-1211 Geneva, Switzerland\\
$^{14}$ASTRON, the Netherlands Institute for Radio Astronomy, Postbus 2, NL-7990 AA Dwingeloo, the Netherlands}
\begin{document}

\date{}

\pubyear{2010}

\maketitle

\label{firstpage}

\begin{abstract}
We present high sensitivity ($\sigma_P \simeq 0.6\,$mJy) polarimetric observations in seven bands, from $2.1$ to $38\,$GHz, of a complete sample of $104$ compact extragalactic radio sources brighter than $200\,$mJy at $20\,$GHz. Polarization measurements in six bands, in the range $5.5-38\,$GHz, for $53$ of these objects were reported by \citet{Galluzzi2017}. We have added new measurements in the same six bands for another 51 sources and measurements at $2.1\,$GHz for the full sample of $104$ sources. Also, the previous measurements at $18$, $24$, $33$ and $38\,$GHz were re-calibrated using the updated model for the flux density absolute calibrator, PKS1934-638, not available for the earlier analysis. The observations, carried out with the Australia Telescope Compact Array (ATCA), achieved a $90\%$ detection rate (at $5\sigma$) in polarization. $89$ of our sources have a counterpart in the $72$ to $231\,$MHz GLEAM survey \citep{HurleyWalker2017}, providing an unparalleled spectral coverage of $2.7$ decades of frequency for these sources. While the total intensity data from $5.5$ to $38\,$GHz could be interpreted in terms of single component emission, a joint analysis of more extended total intensity spectra presented here, and of the polarization spectra, reveals that over $90\%$ of our sources show clear indications of at least two emission components. We interpret this as an evidence of recurrent activity. Our high sensitivity polarimetry has allowed a $5\,\sigma$ detection of the weak circular polarization for $\sim 38\%$ of the dataset, and a deeper estimate of $20\,$GHz polarization source counts than has been possible so far.
\end{abstract}


\section{Introduction}
We have undertaken a long-term observational program to characterize the spectra in total intensity and polarization of a complete sample of extragalactic radio sources over an unprecedented frequency range. The sources were drawn from the \textit{Planck}--ATCA Coeval Observations (PACO) `faint sample' \citep{Bonavera2011}, a complete sample flux-limited at $S_{20\,\rm GHz}\ge 200\,$mJy, extracted from the Australia Telescope Compact Array (ATCA) survey at 20\,GHz \citep[AT20G;][]{Murphy2010}, and observed with the ATCA almost simultaneously with \textit{Planck} observations. The PACO `faint sample' is made up of 159 sources in the South Ecliptic Pole region where \textit{Planck\/}'s scan circles intersect, providing maximal sensitivity. 

In a previous paper \citep{Galluzzi2017} we presented high sensitivity (r.m.s. errors $\sigma_P\simeq 0.6\,$mJy; detection rate of about $91\%$) multi-frequency (six bands, from $5.5$ to $38\,$GHz) polarimetry of a complete sub-sample of $53$ compact extragalactic radio sources at ecliptic latitude $< -75^\circ$, observed in September 2014.

In this work we present new ATCA high sensitivity multi-frequency polarimetric observations of a larger complete sample of `PACO faint' sources, now comprising 104 objects. The frequency coverage was also extended adding polarimetric observations at $2.1\,$GHz. The broader frequency range has allowed a more comprehensive analysis of rotation measures. Moreover, our high sensitivity polarimetry has allowed a $5\,\sigma$ detection of the weak circular polarization for $\sim 38\%$ of data.

The observational determination of total intensity spectra was further extended exploiting the GLEAM (GaLactic and Extra-galactic All-sky Murchison Widefield Array) survey data at $20$ frequencies between $72$ and $231\,$MHz \citep{HurleyWalker2017}, available for $89$ ($\simeq 86\%$) of our sources.

The paper is organised as follows. In Section \ref{sec:obs} we briefly present the observational campaigns. In Section \ref{sec:maths} we describe the data reduction. In Section \ref{sec:dataanaly} we discuss the data analysis and the spectral behaviours in total intensity and polarization. In Section \ref{sec:SouCou} we present source counts in polarization at $\sim20\,$GHz, which is the selection frequency of our sample. Finally, in Section \ref{sec:discusseconcl} we draw our conclusions.

\section{Observations}
\label{sec:obs}

The new observations were carried out in March and April 2016, using the same array configuration (H214) and spectral setup (three sets of $2\times 2$\,GHz CABB -- Compact Array Broadband Backend -- bands centred at 5.5--9, 18--24 and 33--38 GHz) as in the previous campaign held in September 2014, whose results are described by \citet{Galluzzi2017}. The previous sample of $53$ compact extragalactic sources is almost doubled (reaching a total of 104 sources) by adding the `PACO faint' sources at ecliptic latitudes between $-65^\circ$ and $-75^\circ$.

All the 51 additional sources were observed in the six CABB bands. Moreover we re-observed all sources of the September 2014 sample at $5.5$ and $9\,$GHz, while we managed to repeat the observations at $18-24$ and $33-38\,$GHz only for $20\%$ of them. The whole sample of $104$ objects was also observed at $2.1\,$GHz. We obtained three slots in three contiguous days to have the higher frequencies simultaneously observed. The total observing time was of $\simeq 34\,$hr, including overheads and calibration.

In order to achieve the same sensitivity level of previous observations ($\simeq 0.6\,$mJy), we integrated for $1\,$min at $2.1$, $5.5$ and $9\,$GHz, and for $1.5\,$min at the higher frequencies. The effective sensitivity reached in polarization at $2.1\,$GHz is a bit worse than requested, $\simeq 1\,$mJy, due to significant radio-frequency interference (RFI). Weather conditions were good during the whole observing campaign. As done for the September 2014 observations, we considered only data from the $5$ closest antennas in the H214 array configuration, discarding the baselines to the sixth and farthest antenna, since these are the most noisy. The synthesized beam size ranges from $\simeq 90\,$arcsec to $5\,$arcsec in our frequency range.

\section{Data reduction}
\label{sec:maths} 

Data were reduced via the MIRIAD software \citep{Sault1995}. Each frequency band was treated separately, as indicated in the ATCA User's Guide\footnote{www.narrabri.atnf.csiro.au/observing/users\_guide.}. During the data loading, MIRIAD corrects for the time-dependent instrumental xy-phase variation, exploiting the known signal injected from a noise diode, which is mounted in one of the feeds of each antenna.

Our reference for the flux density absolute calibration (at all frequencies) is the source PKS1934-638, a GigaHertz peaked-spectrum (GPS) radio galaxy: it is stable and unpolarized (at least below $30-40\,$GHz)\footnote{It is the only known source with all these characteristics in the Southern sky.}, whose last model \citep[see][]{Sault2003,  Partridge2016} is now loaded into MIRIAD. This introduces a flux density difference with respect to the previous run, so we also re-calibrated the 2014 data. The new flux densities for the $20\,$GHz data are reported in this paper.


Once the calibration tables were derived, all solutions were ingested in the code for flux density extraction. As we did for the September 2014 observations, to better characterize the source spectra, we decided to split each $2\,$GHz-wide frequency band in sub-bands, except for the 2.1\,GHz one that was kept un-split because of the heavy RFI contamination. Each sub-band was calibrated separately. For total intensity, we split each band into $512\,$MHz-wide sub-bands.  For polarized flux densities we split bands in only $2$ sub-bands to limit the $\Delta\nu^{-1/2}$ degradation in sensitivity.

Flux densities were estimated via the MIRIAD task UVFLUX. Our sources are known to exhibit linear polarization \citep[up to $\sim 10\%$;][]{Massardi2008, Massardi2013}, defined by the $Q$ and $U$ Stokes parameters. Observations of the circular polarization of extragalactic radio sources have demonstrated is generally below $0.1-0.2\%$, at least one order of magnitude lower than the linear polarization \citep{Rayner2000}. Hence, the r.m.s. $\sigma_V$ of the retrieved Stokes $V$ parameter is frequently used as a noise estimator.

We achieved a $5\,\sigma$ detection of circular polarization, $V$, in $\sim 38\%$ of the dataset, i.e. $\sim 89\%$ of the objects are detected in Stokes $V$ in at least at one frequency.

Further discussion about the circular polarization is in sub-sect.~\ref{subsec:CirPol}. For only $\sim 15\%$ of detections, the circular to linear polarization ratio is $\ge 20\%$; the mean circular polarization is substantially smaller than our calibration error of the polarized flux density, which is $\simeq 10\%$ \citep{Galluzzi2017}.
Since the contribution of Stokes $V$ is so small, the polarized emission, $P$, can be estimated neglecting the $V$ contribution and adopting $\sigma_V$ as the r.m.s. noise for the Stokes parameters $Q$ and $U$:
\begin{equation}
P=\sqrt{Q^2+U^2-\sigma_V^2}\,.
\label{equ:PolFluDen}
\end{equation}
The $\sigma_V^2$ term removes the noise bias on $P$ \citep[e.g.,][]{WardleKronberg1974}.\footnote{The error associated to the bias correction is negligible and will be ignored in the following.} We find that ignoring the $\sigma_V$ term in eq.~(\ref{equ:PolFluDen}) results in a mean error of 0.01\%.

The polarization angle $\phi$ and fraction $m$ (usually in terms of a percentage) are:
\begin{eqnarray}
\phi&=&\frac{1}{2}\arctan{\left(\frac{U}{Q}\right)},\\
m&=&100\cdot P/I,
\end{eqnarray}
where the Stokes $I$ is the total intensity flux density.
The errors in total intensity, linear polarization flux density and position angle were computed as in \citet{Galluzzi2017}, i.e. adopting calibration errors of $2.5\%$ for $I$ and of a conservative $10.0\%$ for the polarization fraction, $P$, for data between $5.5$ and $38\,$GHz. At $2.1\,$GHz, due to the aforementioned rfi problems, we use a $5\%$ in $I$ and a $12.5\%$ in $P$ as calibration errors. Under the assumption of equal calibration errors for $Q$ and $U$, \citet{Galluzzi2017} reported a $\simeq 3^\circ$ calibration error in the polarization position angle ($3.75^\circ$ at $2.1\,$GHz). For circular polarization  we again assumed a $10\%$ ($12.5\%$ at $2.1\,$GHz) calibration error (i.e. a factor $\simeq \sqrt{2}$ larger than the calibration errors associated to $Q$ and $U$). We note however that, due to the weakness of the signal and the corresponding lack of good calibrators, the calibration error for $V$ is very difficult to estimate.

\section{Data analysis}
\label{sec:dataanaly}
We adopt a $5\sigma$ level for detections in polarization. The median error is $\simeq 0.6\,$mJy. We reach a detection rate of $\simeq 90\%$ for all the sources at all frequencies from $5.5$ to $38\,$GHz. The number of detections is nearly uniform across the observed frequencies ($99$ sources detected at $5.5\,$GHz and $94$ at $38\,$GHz). Following \citet[][their Fig.~1]{Galluzzi2017} we checked the level of intra-band depolarization in this frequency range, by subdividing each $2\,$GHz-wide band into $1\,$GHz-wide sub-bands. No systematic differences were found with respect to the previous assessment. At $2.1\,$GHz, due to the impact of RFI, we cannot proceed with this check, and the detection rate decreases to $\simeq 86\%$. Three of our $107$ observations include the extended source Pictor A. These observations were discarded for the following analysis that therefore deals with $104$ compact objects.

\subsection{Fit procedures}
\label{subsec:fit}

To fit the spectra we used the same functional forms (a double power-law and a triple power-law) of \citet{Galluzzi2017}, and adopted similar criteria about the minimum number of observations to properly constrain the fit parameters. Given the small fraction (less than $10\%$) of non-detections we do not use upper limits when performing the spectral fitting. About $85\%$ of the spectra could be successfully fitted in this way. In only three cases (AT20GJ041239-833521, AT20GJ054641-641522, AT20GJ062524-602030), we do not have detections in polarization at enough frequencies to get a proper fit.

Similarly to what was found for the earlier sample, most ($68\%$) of our source spectra could be fitted with a double power-law down-turning at high frequencies. An upturning double power-law was required in $15$ cases, and a triple power-law in $20$ cases. The median values of the reduced $\chi^2$ are $1.12$ and $1.89$ for Stokes $I$ and $P$, respectively. The spectra for all the sources are presented in Figure~\ref{fig:Spettri1}. The fitting curves and, when available, the previous PACO best epoch ($2009-2010$) observations in total intensity, and the AT20G best epoch (2004-2008) observations in total intensity and in polarization are also presented. In the lower part of each panel we show the polarization fractions (both linear and circular, when detected), followed by the polarization position angles at the different frequencies.

\begin{figure*}
\vspace{-2.5cm}
\centering
\includegraphics[scale=1.0]{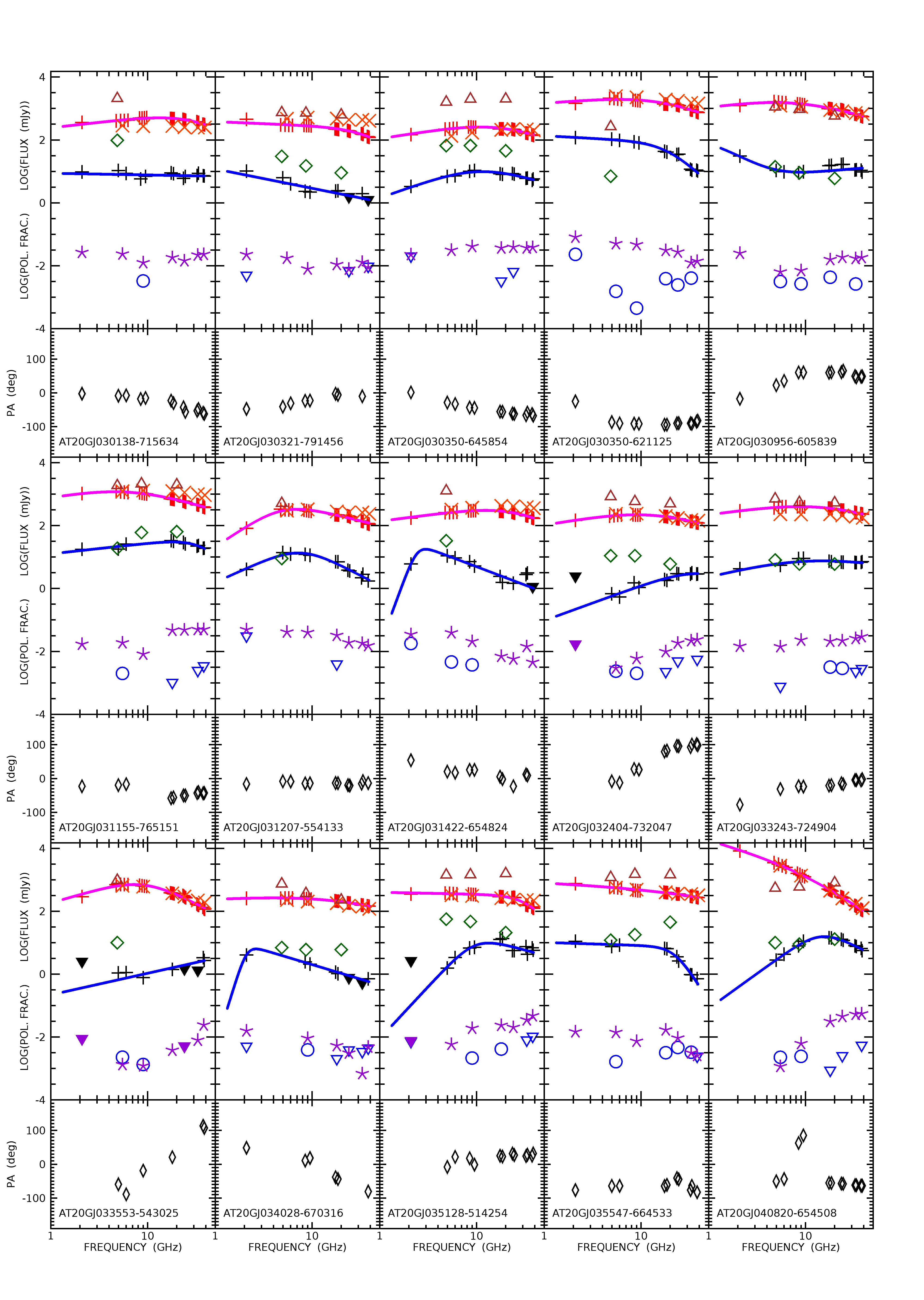}
\vspace{-0.85cm}
\caption{Spectra in total intensity and polarization, polarization fraction and polarization angle for the $107$ objects of the faint PACO sample, observed in the September 2014 and March-April 2016 campaigns. The error bars are not displayed since they are smaller than the symbols. {\bf Total intensity:} red pluses indicate our observations and the solid magenta lines show the fitting curves. The orange crosses show the median PACO flux densities (July 2009-August 2010) while the brown triangles represent the AT20G observations (best epoch in 2004-2008). {\bf Polarization (flux density):} black pluses refer our observations. Upper limits are shown as black filled downwards triangles. The solid blue lines indicate the best fit curves. The AT20G observations (best epoch in 2004-2008) are represented by green diamonds. {\it (Continued...)}}
\label{fig:Spettri1}
\end{figure*}
\addtocounter{figure}{-1}
\begin{figure*}
\vspace{-2.5cm}
\centering
\includegraphics[scale=1.0]{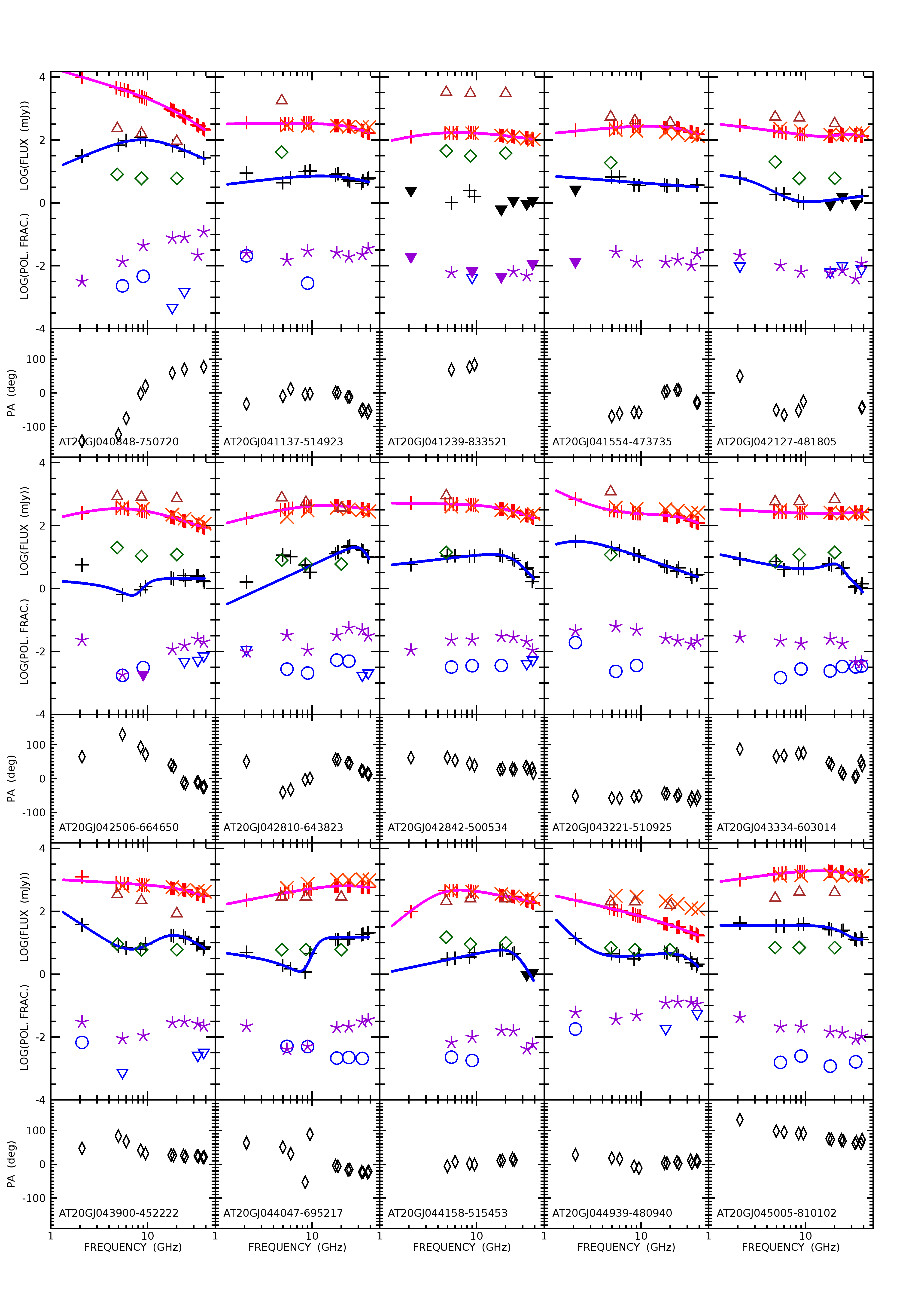}
\caption{{\it (Continued.)} Other quantities available only for the September 2014 and March-April 2016 campaigns: {\bf linear polarization fractions:} purple asterisks with upper limits shown as downwards pointing purple filled triangles; {\bf circular polarization fraction:} violet circles and downward triangles for upper limits. {\bf Polarization angle (PA):} black diamonds.}
\label{fig:Spettri2}
\end{figure*}
\addtocounter{figure}{-1}
\begin{figure*}
\vspace{-2.5cm}
\centering
\includegraphics[scale=1.0]{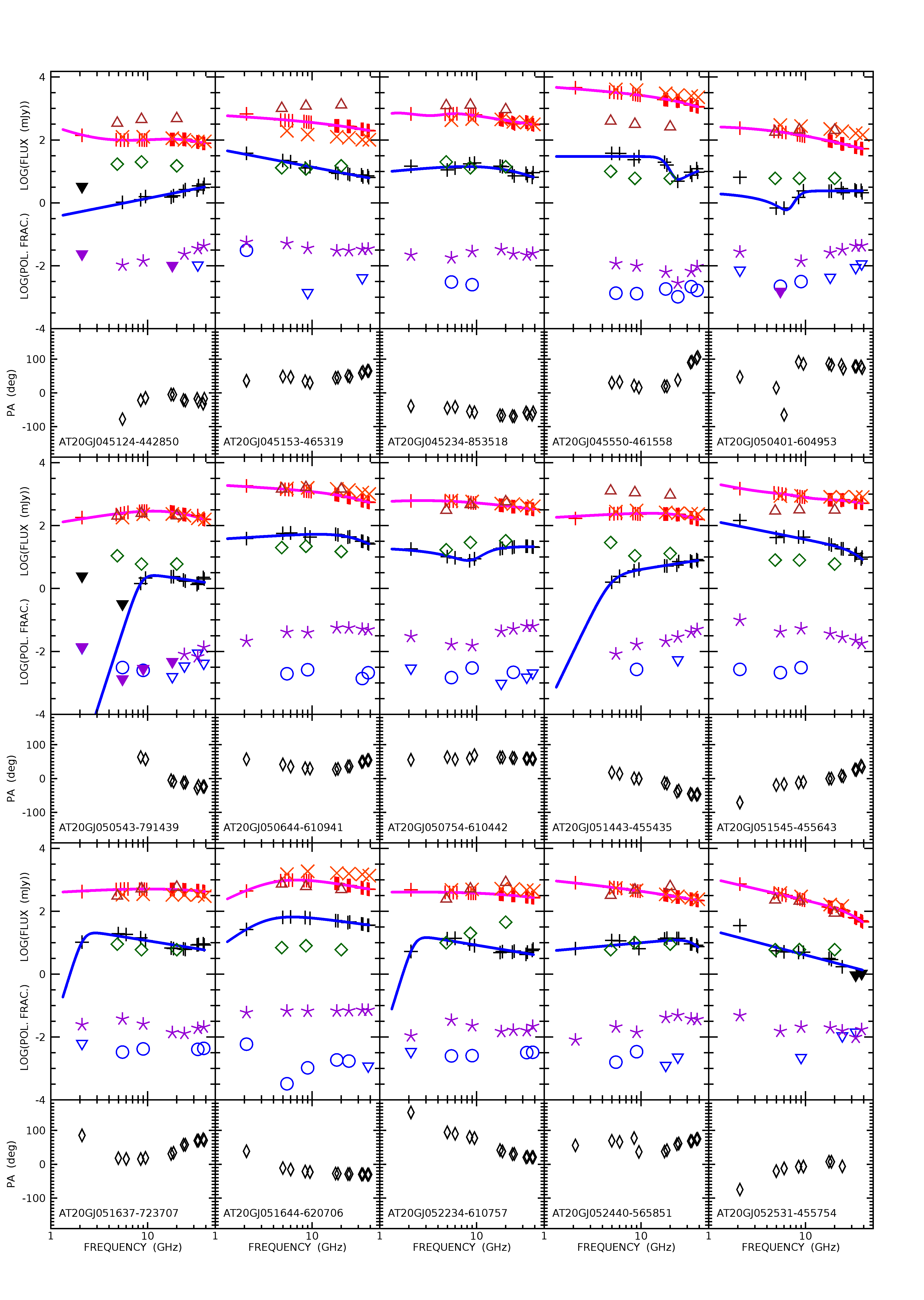}
\caption{Continued.}
\label{fig:Spettri3}
\end{figure*}
\addtocounter{figure}{-1}
\begin{figure*}
\vspace{-2.5cm}
\centering
\includegraphics[scale=1.0]{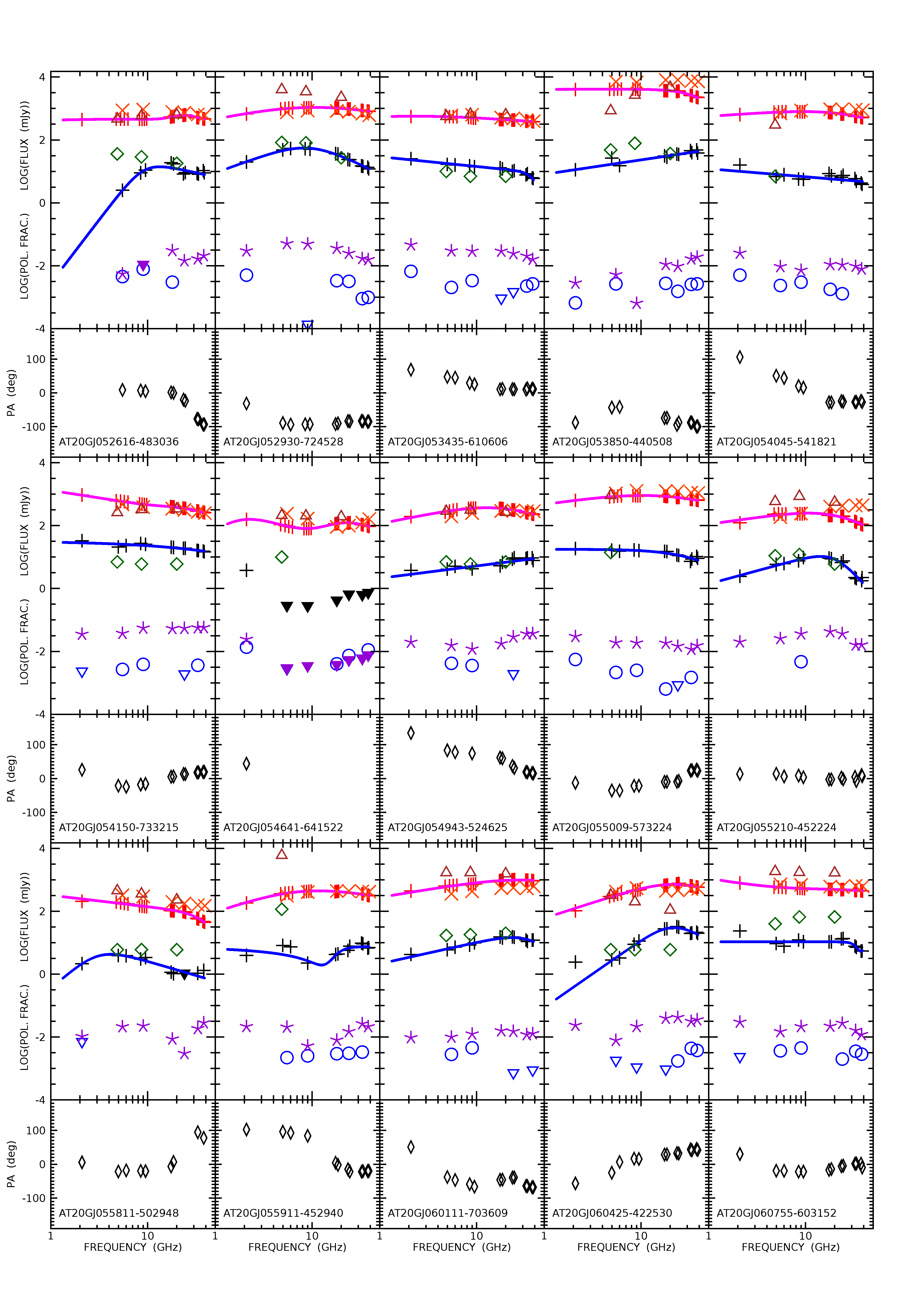}
\caption{Continued.}
\label{fig:Spettri4}
\end{figure*}
\addtocounter{figure}{-1}
\begin{figure*}
\vspace{-2.5cm}
\centering
\includegraphics[scale=1.0]{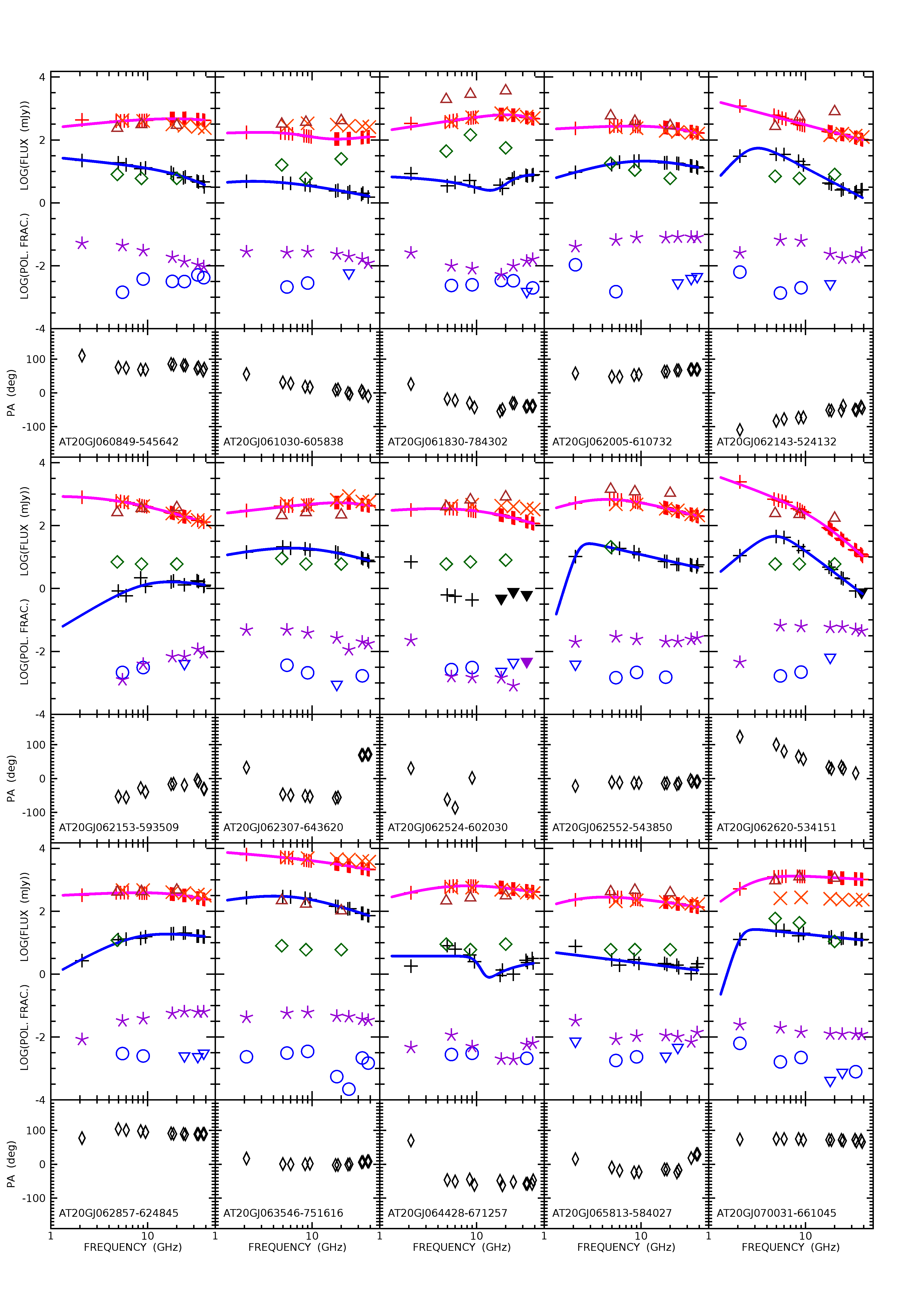}
\caption{Continued.}
\label{fig:Spettri5}
\end{figure*}
\addtocounter{figure}{-1}
\begin{figure*}
\vspace{-2.5cm}
\centering
\includegraphics[scale=1.0]{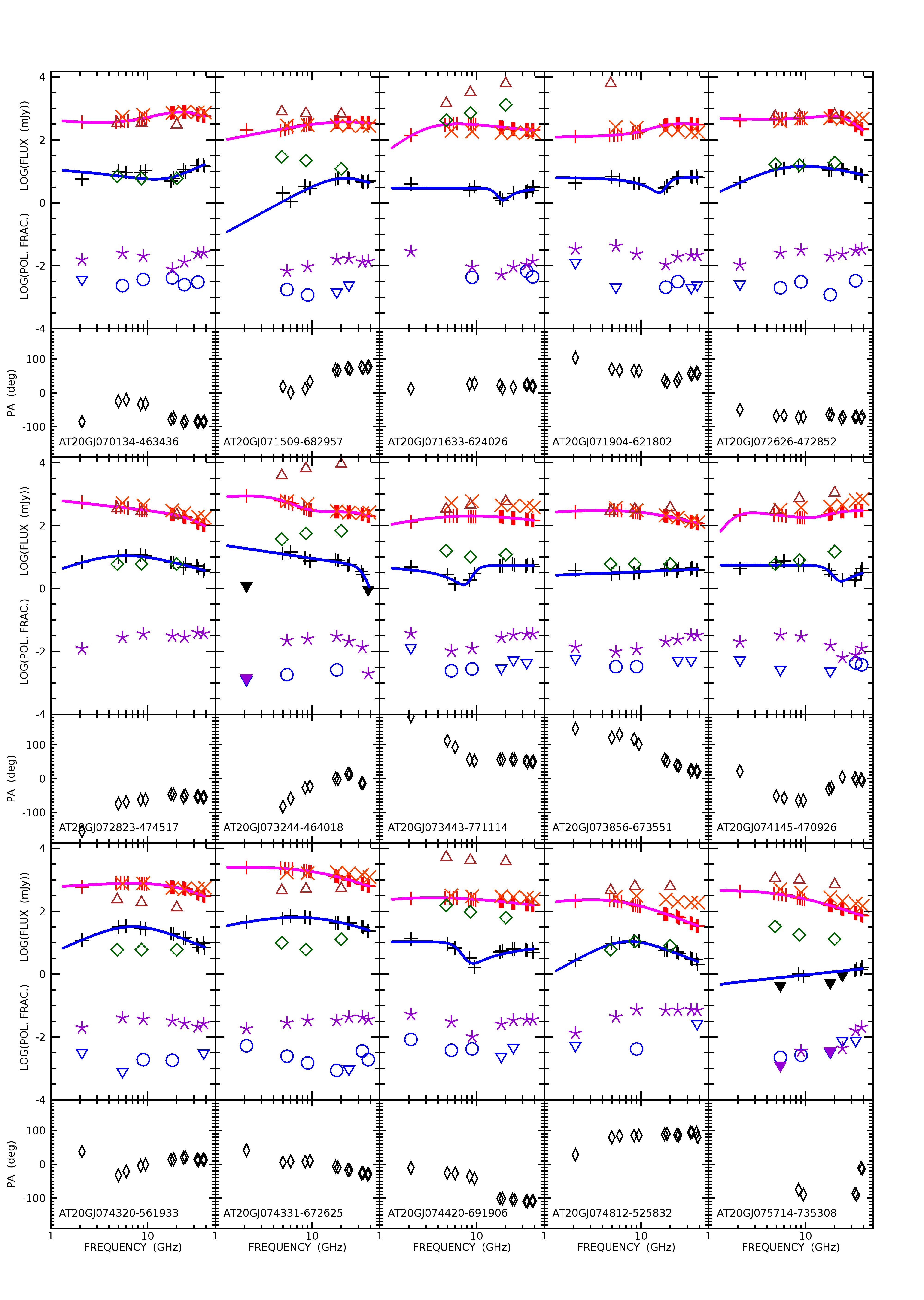}
\caption{Continued.}
\label{fig:Spettri6}
\end{figure*}
\addtocounter{figure}{-1}
\begin{figure*}
\vspace{-2.5cm}
\centering
\includegraphics[scale=1.0]{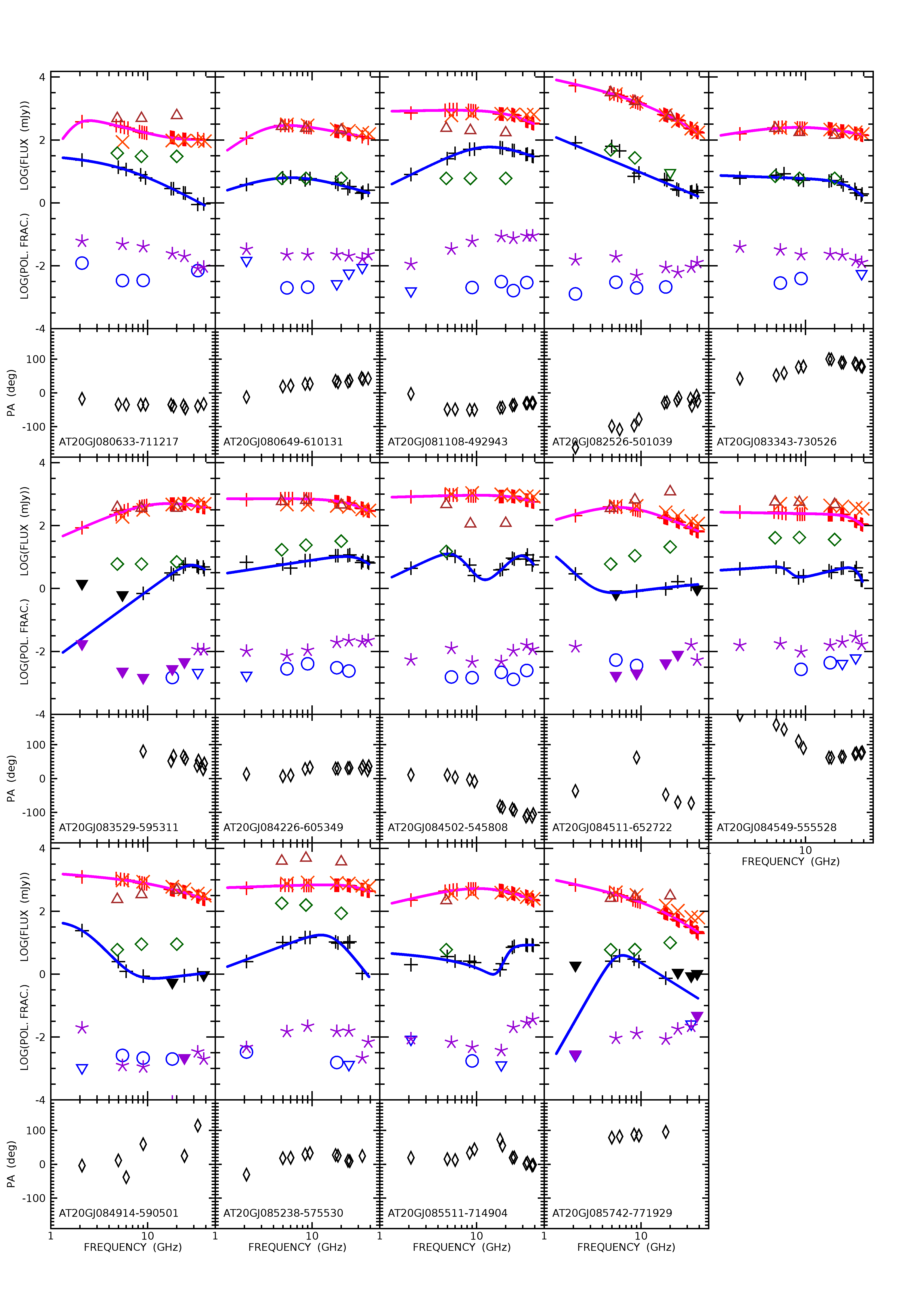}
\caption{Continued.}
\label{fig:Spettri7}
\end{figure*}
\begin{table}
\caption{Distribution of sources per spectral type in total intensity and in polarization. The row `NA' refers to the three objects classified in total intensity but missing a spectral fit in polarization. The last row reports the total for a given spectral class in total intensity, while the last column does the same in polarization.}
\label{tab:Matnumsou}
\vspace{0.3cm}
\begin{tabular}{lccccc||c}
\hline
Tot. Int. $\rightarrow$&(In)&(Pe)&(F)&(S)&(U)&\\
Pol. Int. $\downarrow$ & & & & & &\\
\hline
(In) & 0 & 3 & 0 & 1 & 0 & 4\\
(Pe) & 0 & 24 & 4 & 20& 0 & 48\\
(F) & 0 & 5 & 4 & 4 & 0 & 13\\
(S) & 0 & 5 & 8 & 7& 0 & 20\\
(U) & 0 & 8 & 5 & 3 & 0 & 16\\
(NA)& 0 & 1 & 1&  1 & 0 & 3\\
\hline
\hline
&0&46&22&36&0&\\
\hline
\end{tabular}
\end{table}

\subsection{Spectral properties of the sample}
\label{subsec:speprosam}
%
The spectral index $\alpha_{\nu_1}^{\nu_2}$ between the frequencies $\nu_1$ and $\nu_2$ is defined as:
\begin{equation}
\alpha_{\nu_1}^{\nu_2}=\frac{\log{\left (S(\nu_2)/S(\nu_1)\right )}}{\log{\left (\nu_2/\nu_1\right )}} \,,
\label{equ:SpeInd}
\end{equation}
where $S(\nu_1)$ and $S(\nu_2)$ are the flux densities  associated to the two frequencies. With respect to the previous work \citep{Galluzzi2017} we simply add the $2.5\,$GHz to the reference frequencies $5.5$, $10$, $18$, $28$ and $38\,$GHz in order to preserve the equal spacing in logarithmic scale. Then, we proceed as usual for the spectral classification, taking into account $\alpha_{2.5}^{5.5}$ and $\alpha_{28}^{38}$ and again distinguishing in flat- (F), steep- (S), peaked- (Pe), inverted- (In) and upturning-spectrum (U) object. 

\begin{table*}
\caption{First, second (median), and third quartiles of spectral indices in total intensity and in polarization for different frequency ranges. We give values for the full sample and for the three main spectral classes, as classified in total intensity.}
\label{tab:Medvalspeind2}
\vspace{0.3cm}
\begin{tabular}{|l|c|c|c|}
\hline
Tot. Int. & $2.5-5.5$ & $5.5-10$ & $10-18\,$GHz\\
\hline
\begin{tabular}{l}Quart.\\
\hline
All \\
Steep \\
Peaked \\
Flat
\end{tabular} &
\begin{tabular}
{lcr}{\scriptsize Q1}  & Q2 & {\scriptsize Q3}\\
\hline
{\scriptsize -0.29} & $-0.02$ & {\scriptsize 0.31}\\
{\scriptsize -0.64} & $-0.33$ & {\scriptsize -0.13}\\
{\scriptsize 0.10} & $0.32$ & {\scriptsize 0.54} \\
{\scriptsize -0.16} & $-0.01$ & {\scriptsize 0.27}
\end{tabular}
&
\begin{tabular}
{lcr} {\scriptsize Q1} & Q2 & {\scriptsize Q3} \\
\hline
{\scriptsize -0.35} & $-0.11$ & {\scriptsize 0.09} \\
{\scriptsize -0.67} & $-0.37$ & {\scriptsize -0.22} \\
{\scriptsize -0.04} & $0.06$ & {\scriptsize 0.27} \\
{\scriptsize -0.25} & $-0.12$ & {\scriptsize 0.06}
\end{tabular}
&
\begin{tabular}
{lcr} {\scriptsize Q1} & Q2 & {\scriptsize Q3} \\
\hline
{\scriptsize -0.46} & $-0.24$ & {\scriptsize -0.08} \\
{\scriptsize -0.80} & $-0.46$ & {\scriptsize -0.30} \\
{\scriptsize -0.30} & $-0.14$ & {\scriptsize 0.01} \\
{\scriptsize -0.26} & $-0.14$ & {\scriptsize -0.05}
\end{tabular}

\\
\hline
\end{tabular}

\vspace{0.3cm}

\begin{tabular}{|l|c|c|}
\hline
Tot. Int. & $18-28$ & $28-38\,$GHz\\
\hline
\begin{tabular}{l}Quart.\\
\hline
All \\
Steep \\
Peaked \\
Flat
\end{tabular} &
\begin{tabular} 
{lcr} {\scriptsize Q1} & Q2 & {\scriptsize Q3} \\
\hline
{\scriptsize -0.75} & $-0.46$ & {\scriptsize -0.27} \\
{\scriptsize -0.91} & $-0.76$ & {\scriptsize -0.56} \\
{\scriptsize -0.56} & $-0.42$ & {\scriptsize -0.28} \\
{\scriptsize -0.33} & $-0.26$ & {\scriptsize 0.09}
\end{tabular}
&
\begin {tabular}
{lcr} {\scriptsize Q1} & Q2 & {\scriptsize Q3} \\
\hline
{\scriptsize -1.00} & $-0.75$ & {\scriptsize -0.44} \\
{\scriptsize -1.56} & $-1.02$ & {\scriptsize -0.81} \\
{\scriptsize -0.85} & $-0.74$ & {\scriptsize -0.56} \\
{\scriptsize -0.42} & $-0.34$ & {\scriptsize -0.20}
\end{tabular}
\\
\hline
\end{tabular}

\vspace{0.5cm}

\begin{tabular} {|l|c|c|c|}

\hline
Pol. Int. & $2.5-5.5$ & $5.5-10$ & $10-18\,$GHz\\
\hline
\begin{tabular} {l} Quart. \\
\hline
All \\
Steep \\
Peaked\\
Flat
\end{tabular}
&
\begin{tabular}
{lcr} {\scriptsize Q1} & Q2 & {\scriptsize Q3} \\
\hline
{\scriptsize -0.28} & $0.15$ & {\scriptsize 0.85} \\
{\scriptsize -0.46} & $0.33$ & {\scriptsize 0.61} \\
{\scriptsize -0.06} & $0.49$ & {\scriptsize 1.04} \\
{\scriptsize -0.57} & $-0.21$ & {\scriptsize -0.06}
\end{tabular}
&
\begin{tabular}
{lcr} {\scriptsize Q1} & Q2 & {\scriptsize Q3} \\
\hline
{\scriptsize -0.43} & $-0.06$ & {\scriptsize 0.38} \\
{\scriptsize -0.59} & $-0.06$ & {\scriptsize 0.33} \\
{\scriptsize -0.19} & $-0.01$ & {\scriptsize 0.71} \\
{\scriptsize -0.54} & $-0.29$ & {\scriptsize -0.06}
\end{tabular}
&
\begin{tabular}
{lcr}{\scriptsize Q1} & Q2 & {\scriptsize Q3} \\
\hline
{\scriptsize -0.61} & $-0.15$ & {\scriptsize 0.34} \\
{\scriptsize -0.74} & $-0.24 $ & {\scriptsize 0.29} \\
{\scriptsize -0.35} & $-0.06$ & {\scriptsize 0.36} \\
{\scriptsize -0.59} & $-0.33$ & {\scriptsize 0.54}
\end{tabular}

\\
\hline
\end{tabular}

\vspace{0.3cm}

\begin{tabular} {|l|c|c|}
\hline
Pol. Int. & $18-28$ & $28-38\,$GHz\\
\hline
\begin{tabular} {l} Quart. \\
\hline
All \\
Steep \\
Peaked \\
Flat
\end{tabular}
&
\begin{tabular}
{lcr} {\scriptsize Q1} & Q2 & {\scriptsize Q3} \\
\hline
{\scriptsize -0.98} & $-0.53$ & {\scriptsize 0.02} \\
{\scriptsize -1.00} & $-0.76$ & {\scriptsize -0.10} \\
{\scriptsize -0.80} & $-0.32$ & {\scriptsize 0.31} \\
{\scriptsize -1.01} & $-0.54$ & {\scriptsize 0.14}
\end{tabular} &
\begin{tabular}
{lcr}{\scriptsize Q1} & Q2 & {\scriptsize Q3} \\
\hline {\scriptsize -1.44} & $-0.80$ & {\scriptsize -0.03} \\
{\scriptsize -1.47} & $-0.92$ & {\scriptsize -0.37} \\
{\scriptsize -1.21} & $-0.73$ & {\scriptsize -0.23} \\
{\scriptsize -1.61} & $-0.68$ & {\scriptsize 0.04}
\end{tabular}
\\
\hline
\end{tabular}
\end{table*}


%

\begin{figure*}
\includegraphics[width=\columnwidth]{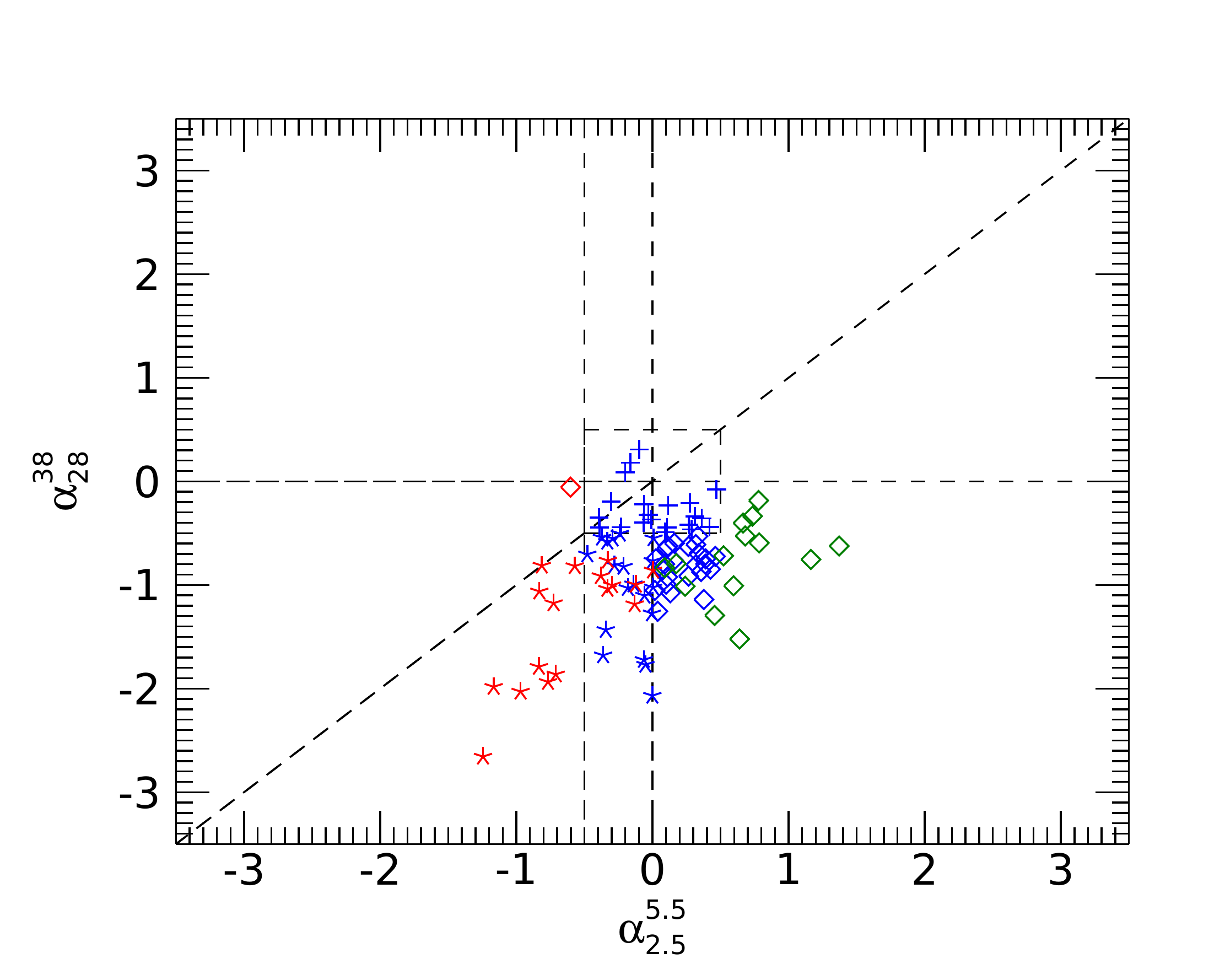}
\includegraphics[width=\columnwidth]{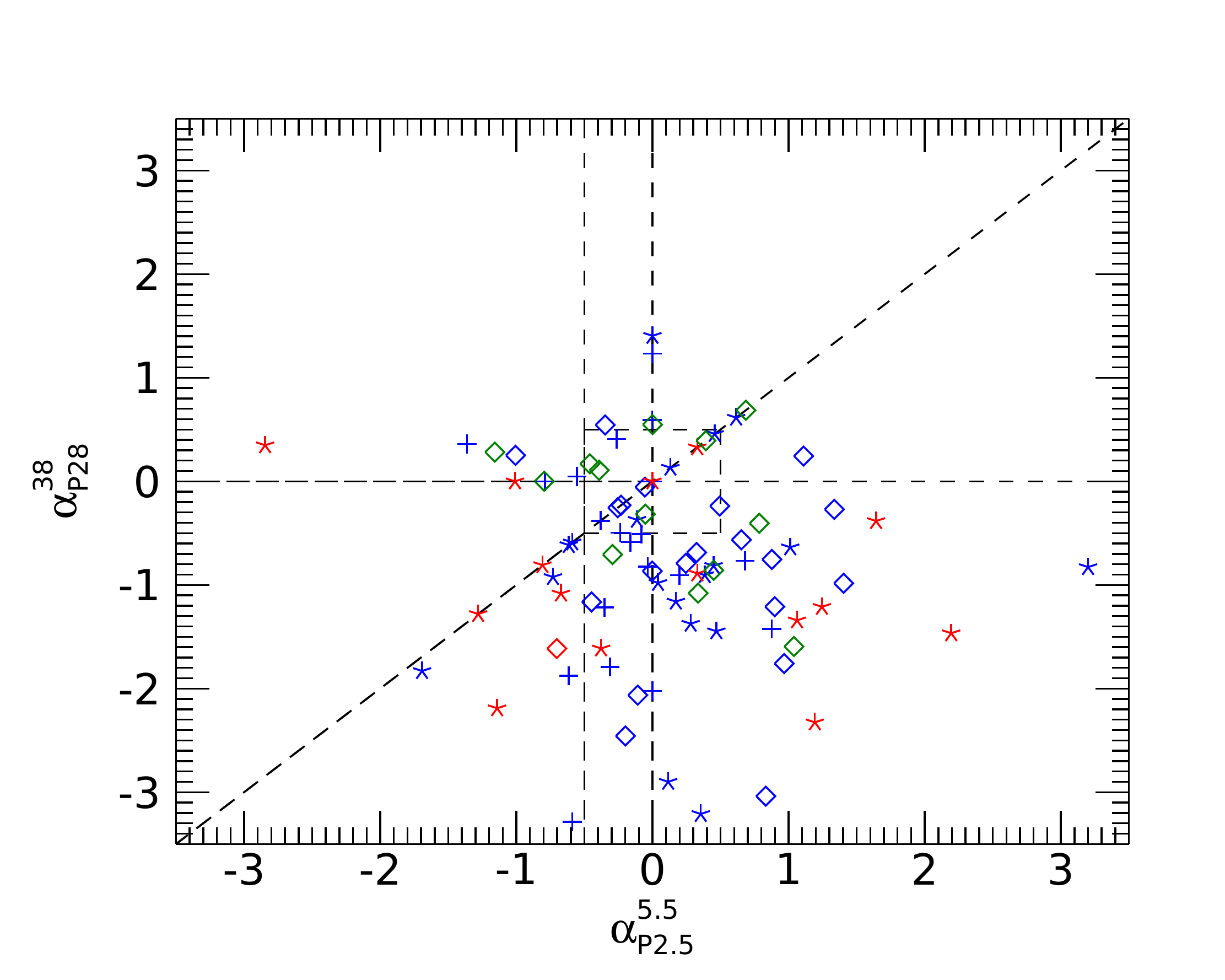}
\caption{Radio colour-colour diagrams for (from left to right) total intensity and polarized flux density. Symbols identify the spectral type in total intensity: pluses for flat-spectrum, asterisks for steep-spectrum, diamonds for peaked-spectrum. Colours refer to the spectral shape between $5.5$ and $18\,$GHz: red for steep-spectrum, blue for flat-spectrum, and green for peaked-spectrum sources.}
\label{fig:ColColPlo}
\end{figure*}

We populate the Table~\ref{tab:Matnumsou} with the outcome of the classification performed in total intensity and polarization, while we report the quartiles of distributions of spectral indices in Table~\ref{tab:Medvalspeind2}. The basic conclusions of our analysis of the earlier, small sample are confirmed: less than $40\%$ of sources have the same spectral behaviour in total intensity and in polarization, and high- and low-frequency spectral indices are essentially uncorrelated, as shown in Fig.~\ref{fig:ColColPlo}.
The most populated entries of Table~\ref{tab:Matnumsou} are sources peaking both in total intensity and polarization and sources which are steep-spectrum in total intensity but have a spectral peak in polarization. This change in spectral shape toward a peaked- or even an upturning-spectrum in polarization might be the sign of Faraday depolarization which typically lowers the polarization signal at lower frequencies.

\citet{Galluzzi2017} pointed out that past high frequency flux density measurements may suffer from the low accuracy of the model for the primary calibrator. In fact, they found that the mean high-frequency spectral index of `PACO faint' sources in total intensity reported by \citet{Bonavera2011} was flatter by $\delta\alpha \sim 0.3$. In this work we have taken into account the new model for the primary calibrator (PKS 1934-638) now encoded into MIRIAD. Since the model was not implemented yet, \citet{Galluzzi2017} applied a-posteriori corrections.

The differences with the results by \citet{Galluzzi2017} are relatively small. We confirm that the high-frequency spectral indices in total intensity and in polarization steepen at high frequencies and are essentially uncorrelated, although the mean values ($\alpha_{28}^{38}\simeq -0.75$ and $\alpha_{{\rm P }28}^{38}\simeq -0.80$, respectively) are less steep and closer to each other than found by \citet{Galluzzi2017}. The distribution of sources among the different spectral types is also very similar; the biggest difference is in the fraction of objects classified as flat-spectrum in total intensity that increases from $\simeq 4\%$ to $\simeq 21\%$.

To extend the spectral coverage we have exploited the information provided by the GLEAM (GaLactic and Extragalactic All-sky Murchison Widefield Array) survey at $20$ frequencies between $72$ and $231\,$MHz \citep{HurleyWalker2017}. The spatial resolution is $\simeq 2\,$arcmin at $200\,$MHz, similar to the $\sim 90\,$arcsec resolution of our $2.1\,$GHz observations. We have $89$ matching sources ($\simeq 86\%$ of our sample) in the GLEAM survey. For these sources we have the unparalleled coverage of $2.7$ decades in frequency. Since the GLEAM survey covers all the sky south of $+30^\circ$ in declination with a mean sensitivity of $\sim 10\,$mJy, and our sample is located between $-86^\circ$ and $-42^\circ$, we can associate an upper limit of $50\,$mJy (at $5\sigma$) to those sources without a GLEAM counterpart.

The fitting curves (triple power-laws), although not always successful, generally show a good consistency between the ATCA and GLEAM measurements (cf. Fig.~\ref{fig:Spettri1}). But while in the range $5.5\,\hbox{GHz}- 38\,\hbox{GHz}$ the spectra are consistent with a single emitting region \citep{Galluzzi2017}, the GLEAM flux densities are clearly above the extrapolations from higher frequencies in $\sim 40\%$ of the cases, strongly suggesting the presence of at least another, generally steeper, component. The joint analysis with polarization data suggests even more complex structures (cf. \citet{Farnes2014}, see sub-section~\ref{sect:LP}).

\begin{figure*}
\includegraphics[scale=0.60]{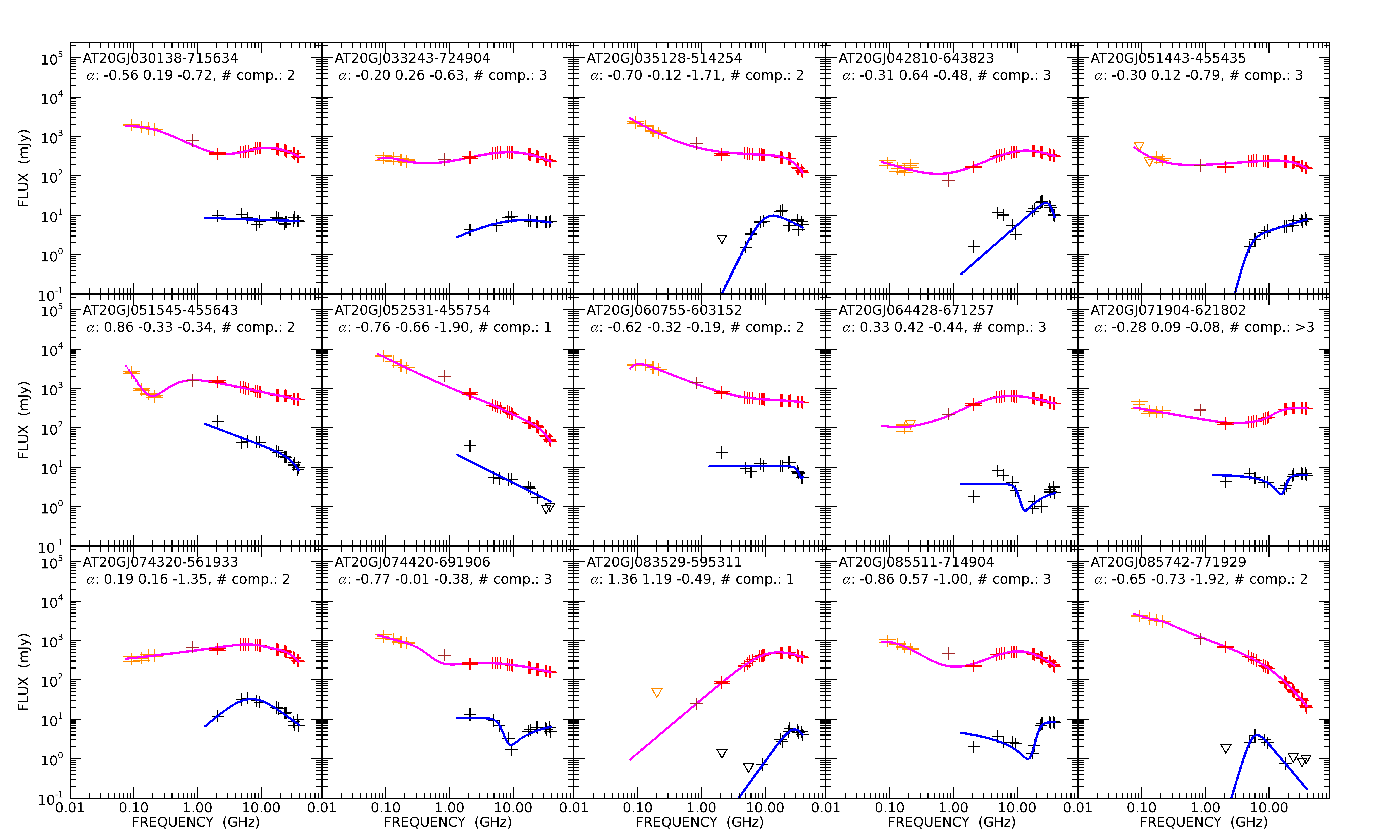}
\caption{Spectra in total intensity and polarization taking into account GLEAM observations between $72\,$ and $231\,$MHz (orange pluses). Our observations in total intensity are the red pluses, while black pluses and downward triangles are for polarization. The fits in total intensity and polarization are given as red and black solid curves, respectively. We report the SUMSS (Sydney University Molongo Sky Survey, \citet{Mauch2003}) flux density at $843\,$MHz just for comparison, but we do not use it in the fit procedures. For each object, after its name, spectral indices (computed in total intensity) $\alpha_{0.2}^{0.4}$, $\alpha_{2.5}^{5.5}$ and $\alpha_{28}^{38}$ are provided. At the end of the title of each plot there is the spectral classification in terms of estimated synchrotron components (see~\ref{sect:LP}).}
\label{fig:Spettri_wGLEAM}
\end{figure*}

\subsection{Linear polarization fraction}\label{sect:LP}

\citet{Galluzzi2017} did not find any systematic variation of the mean polarization fraction with either flux density or frequency, down to $\simeq 5\,$GHz, in agreement with the results by \citet{Massardi2013}. A similar conclusion was reached by \citet{Battye2011}, who however had measurements only down to 8.4\,GHz. On the other hand, claims of a systematic decrease of the polarization fraction with decreasing frequency were made by \citet{Agudo2010,Agudo2014} and \citet{Sajina2011}, suggesting that Faraday depolarization may work up to $\simeq 10\,$GHz or that the magnetic field is more ordered at high frequencies \citep{Tucci2004}. However the conclusions by \citet{Agudo2010,Agudo2014} and \citet{Sajina2011} may be biased towards greater polarization fractions by not having taken into account non-detections \citep{TucciToffolatti2012}.

Our new observations allowed us to extend the spectra in polarized intensity down to $2\,$GHz and to have a factor of $2$ larger sample. As for \citet{Galluzzi2017}, our high detection rate (over 90\%) safeguards against any selection bias. Although the polarization fraction declines for several sources drops at lowest frequency (cf. Fig.~\ref{fig:Spettri1}), there is no statistical evidence of a decrease of the mean value for the whole sample or for its sub-samples. However, as discussed below, such apparent uniformity may hide a more complex situation. The steep-spectrum objects ($36$) indeed show a slight trend, but comparing to the distributions of polarization fraction at $2.1\,$ and $38\,$GHz, the rejection of the null hypothesis reaches the $\simeq 2\,\sigma$ level. The sample of flat-spectrum objects ($22$ objects in total) seems to reveal an opposite trend, but also in this case the significance is less than $3\sigma$.

The spectra of the polarization fraction are less smooth than the total intensity spectra. Only about $15\%$ of the sources have an approximately constant polarization fraction over the full frequency range. Five sources with smooth total intensity spectra above $2\,$GHz have double peaked fractional polarization, suggesting at least two emission components, seeing different screens. The polarization fraction of $\sim 15\%$ of the sources has an upturn at 2 GHz, where the emission components seen in the GLEAM data may yield a substantial contribution. The polarized flux from these components can drown out the decrease of the polarization fraction of the higher frequency component, due to Faraday depolarization. The most straightforward interpretation of these results is that the extension (and, correspondingly, the age) of emission components increases with decreasing frequency.

On the whole, a joint inspection of total intensity (including GLEAM measurements between $72\,$MHz and $231\,$MHz) and polarization spectra indicates the presence of at least 2 (sometimes 3) emission components for about $93\%$ of the sources. This is expected for GPS/CSS sources due to their double lobe structure \citep{Tingay2003,Callingham2015}. For about half of these, the clearest indication comes from polarization data. Hence, we reclassify our sample by distinguishing cases in which there is no sign of an additional synchrotron component (we label it `1C') from situations in which there are hints of $2-3$ synchrotron components (`2-3C') or more complicated cases which seems to reveal more than $3$ components in the spectrum. The latter are quite flat sources in total intensity from $70\,$MHz up to $\sim 30\,$GHz, where a steepening typically occurs. Among these $17$ objects ($\simeq 16\%$) $10$ are classified in the flat (F) spectral category, i.e. objects with a flat spectrum in total intensity between $2.1$ and $38\,$GHz.

According to Fig.~\ref{fig:PPolNCVsfreq} we do not have evidences of trends of the linear polarization fraction with the frequency for the full sample and for `1C' sources. `2-3C' sources have a minimum of the polarization fraction at $\simeq 9\,$GHz, consistent with different emission components at lower and higher frequencies. For the $>$3C objects, whose spectra show indications of several overlapping synchrotron components, there is a hint of a \textit{decrease} with increasing frequency \citep[rather than of the increase expected by some authors, see e.g.][]{TucciToffolatti2012} of the polarization fraction: the mean values decline from $\simeq 2.1-2.4\%$ at $\le 5.5\,$GHz to $1.2\%$ at $38\,$GHz. We anticipate here that in sub-section~\ref{sect:CP}, we find these sources to have very large rotation measures (RMs) at mm wavelengths. This could indicate that their high frequency components are characterised by a really dense and/or a magnetised medium that rotates the polarisation angle strongly \citep[cf.][]{Pasetto2016}.

\begin{table*}
\caption{First, second (median) and third quartiles of the polarization fraction at each observed frequency given by the Kaplan-Meier estimator, taking into account the upper limits, for the full sample and for the steep- and peaked-spectrum sources. The last row reports probabilities for the null hypothesis (i.e. the two samples are drawn from the same parent distribution) given by the Kolmogorov-Smirnov test performed on the steep and peaked groups, considering together $5.5$ and $9\,$GHz, the $18-38\,$GHz frequency interval and all the frequencies, respectively.}
\label{tab:PPolVsfreq}
\begin{tabular}{cc}
\hline
Class.& frequencies (GHz)\\
\begin{tabular}{@{}c@{}}
\\
\hline
\\
All\\
Steep\\
Peaked\\
Flat\\
\hline
Prob.
\end{tabular} &
\hspace{-0.3cm}
\begin{tabular} {@{}ccccccc@{}}
2.1 &
\hspace{-0.3cm}
5.5 &
\hspace{-0.3cm}
9 &
\hspace{-0.3cm}
18
\\
\hline
\begin{tabular} 
{lcr} {\scriptsize Q1} & Q2 & {\scriptsize Q3} \\ 
\hline 
{\scriptsize 1.09} & $2.16$ & {\scriptsize 2.98} \\ 
{\scriptsize 0.95} & $1.54$ & {\scriptsize 2.54} \\
{\scriptsize 1.14} & $2.22$ & {\scriptsize 2.75} \\
{\scriptsize 1.73} & $2.79$ & {\scriptsize 3.36} 
\end{tabular} 
&
\hspace{-0.3cm}
\begin{tabular}
{lcr} {\scriptsize Q1} & Q2 & {\scriptsize Q3} \\
\hline
{\scriptsize 0.84} & $1.88$ & {\scriptsize 3.25} \\
{\scriptsize 0.81} & $1.74$ & {\scriptsize 3.23} \\
{\scriptsize 0.67} & $1.71$ & {\scriptsize 3.16} \\
{\scriptsize 1.09} & $1.88$ & {\scriptsize 3.38}
\end{tabular}
&
\hspace{-0.3cm}
\begin{tabular}
{lcr} {\scriptsize Q1} & Q2 & {\scriptsize Q3} \\
\hline
{\scriptsize 0.79} & $1.65$ & {\scriptsize 3.02} \\
{\scriptsize 0.95} & $1.64$ & {\scriptsize 3.69} \\
{\scriptsize 0.64} & $1.54$ & {\scriptsize 2.78} \\
{\scriptsize 1.19} & $2.00$ & {\scriptsize 2.81}
\end{tabular}
&
\hspace{-0.3cm}
\begin{tabular}
{lcr} {\scriptsize Q1} & Q2 & {\scriptsize Q3} \\
\hline
{\scriptsize 0.99} & $2.01$ & {\scriptsize 3.07} \\
{\scriptsize 0.87} & $2.33$ & {\scriptsize 3.13} \\
{\scriptsize 0.85} & $1.75$ & {\scriptsize 3.05} \\
{\scriptsize 1.35} & $1.82$ & {\scriptsize 2.67}
\end{tabular}

\\
\hline

{\scriptsize ($5.5-9\,$GHz)}
&
\hspace{-1cm}
{\scriptsize $0.825$}
&
{\scriptsize ($18-38\,$GHz) }
&
\hspace{-1cm}
{\scriptsize $8.176\cdot 10^{-4}$}
\end{tabular}
\\
\hline
\end{tabular}

\vspace{0.6cm}

\begin{tabular}{cc}
\hline
Class.& frequencies (GHz)\\
\begin{tabular}{@{}c@{}}
\\
\hline
\\
All\\
Steep\\
Peaked\\
Flat\\
\hline
Prob.
\end{tabular} &
\hspace{-0.3cm}
\begin{tabular} {@{}ccccccc@{}}
24 &
\hspace{-0.3cm}
33 &
\hspace{-0.3cm}
38 \\
\hline
\begin{tabular}
{lcr}{\scriptsize Q1} & Q2 & {\scriptsize Q3} \\
\hline
{\scriptsize 1.06} & $1.95$ & {\scriptsize 2.87} \\
{\scriptsize 0.66} & $2.19$ & {\scriptsize 4.28} \\
{\scriptsize 1.26} & $1.84$ & {\scriptsize 2.72} \\
{\scriptsize 0.98} & $1.64$ & {\scriptsize 2.45}
\end{tabular}
&
\hspace{-0.3cm}
\begin{tabular}
{lcr}{\scriptsize Q1} & Q2 & {\scriptsize Q3} \\
\hline
{\scriptsize 1.17} & $1.85$ & {\scriptsize 3.29} \\
{\scriptsize 1.31} & $2.37$ & {\scriptsize 3.84} \\
{\scriptsize 1.25} & $1.75$ & {\scriptsize 2.81} \\
{\scriptsize 0.66} & $1.60$ & {\scriptsize 2.08}
\end{tabular}
&
\hspace{-0.3cm}
\begin{tabular}
{lcr} {\scriptsize Q1} & Q2 & {\scriptsize Q3} \\
\hline
{\scriptsize 1.20} & $2.09$ & {\scriptsize 3.54} \\
{\scriptsize 1.17} & $2.62$ & {\scriptsize 4.00}\\
{\scriptsize 1.31} & $2.11$ & {\scriptsize 3.42} \\
{\scriptsize 1.06} & $1.40$ & {\scriptsize 2.11}
\end{tabular}
\\
\hline
{\scriptsize (All freqs.)} &
\hspace{-1cm}
{\scriptsize $0.011$}
\end{tabular}\\
\hline
\end{tabular}
\end{table*}

\begin{figure}
\includegraphics[width=\columnwidth]{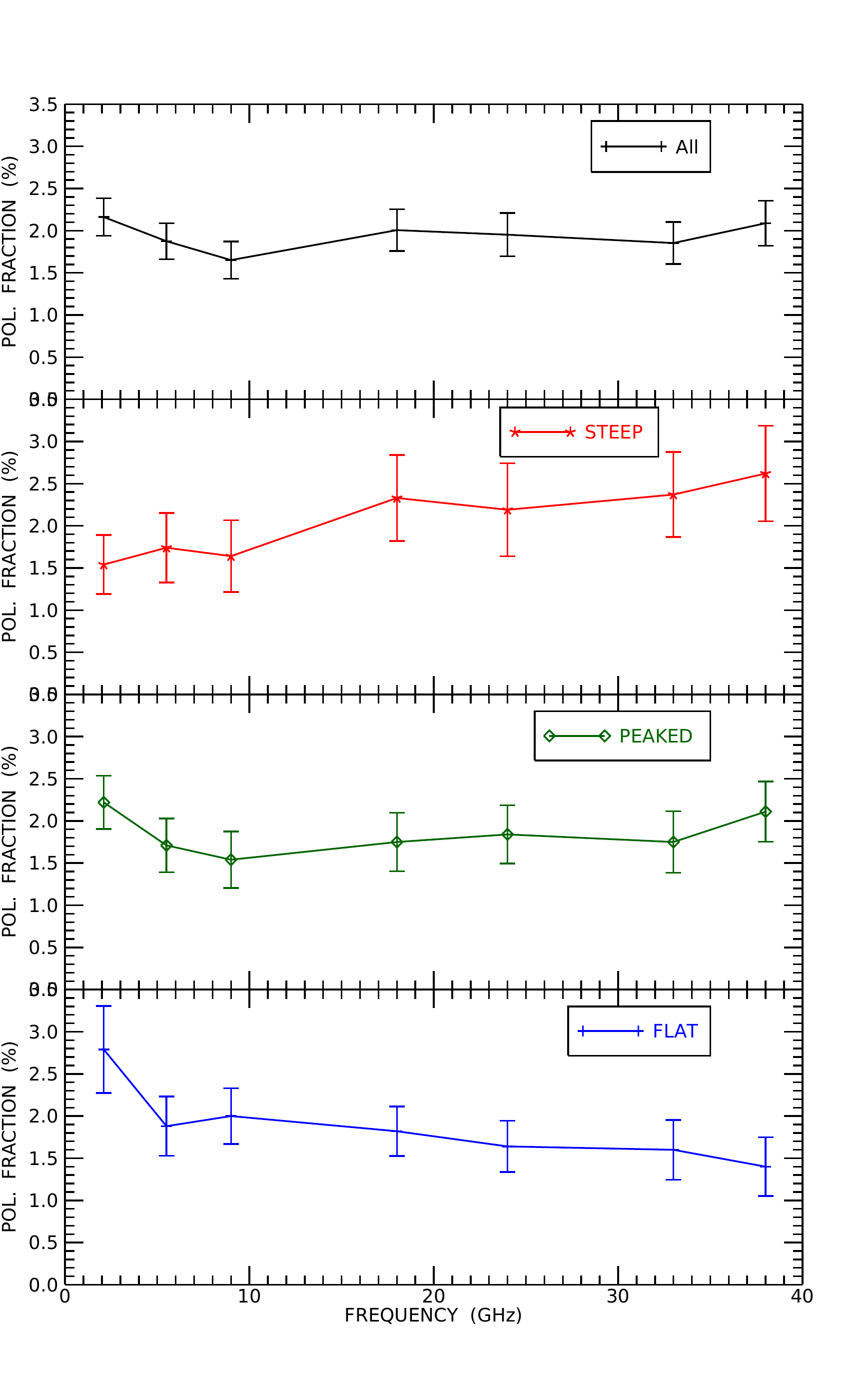}
\caption{Median polarization fraction behaviour with frequency (at $2.1$, $5.5$, $9$, $18$, $24$, $33$ and $38\,$GHz) for all the sources (black), for steep sources (red), for peaked (green) and flat ones (blue). The errors on median values are given by $1.253\times{\rm rms}/\sqrt{N}$, where rms is the standard deviation around the mean and $N$ is the number of the data (at a given frequency) for a given class of objects (cf.~Arkin \& Colton 1970).}
\label{fig:PPolVsfreq}
\end{figure}

\begin{figure}
\includegraphics[width=\columnwidth]{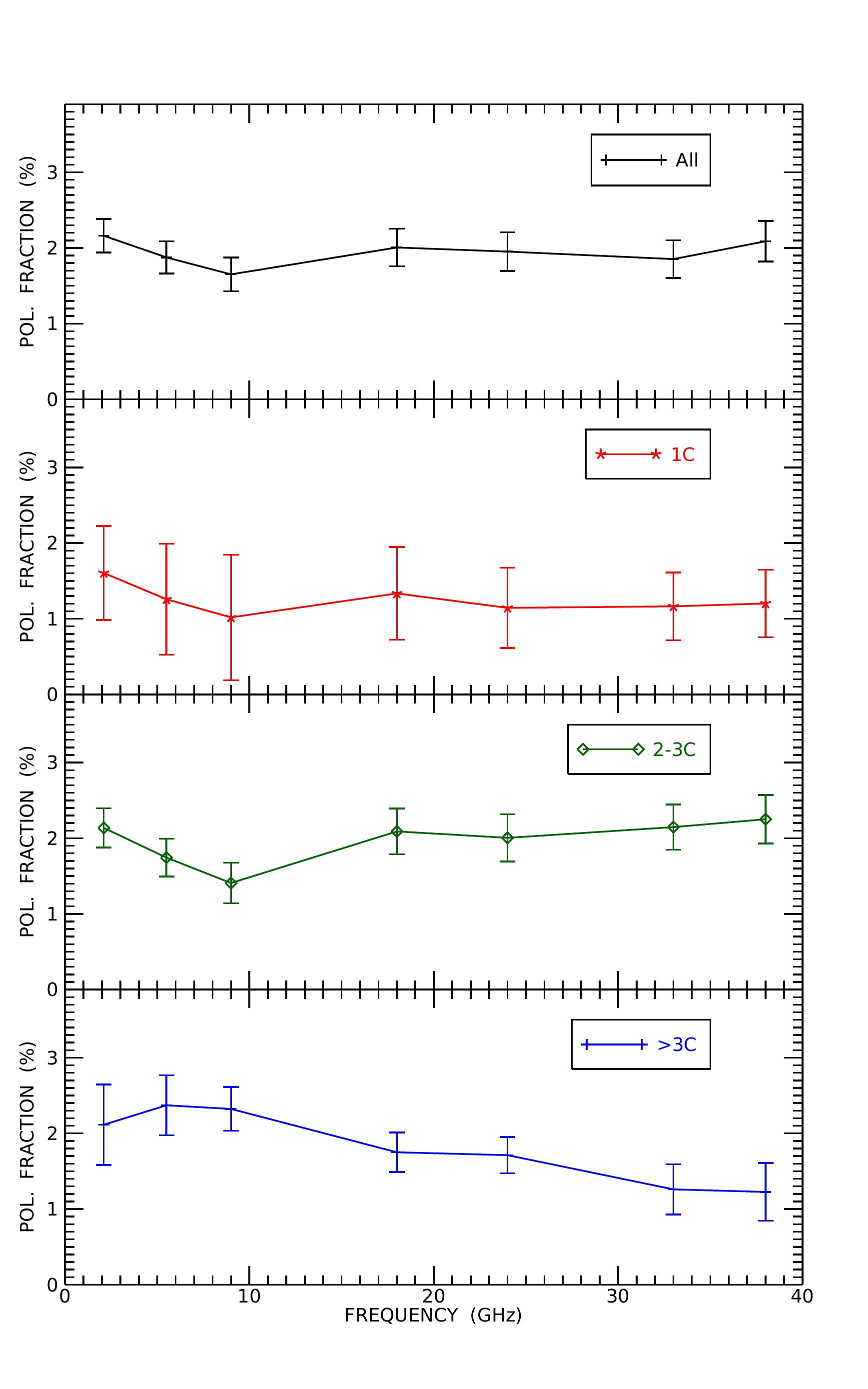}
\caption{Median polarization fraction at the observation frequencies ($2.1$, $5.5$, $9$, $18$, $24$, $33$ and $38\,$GHz) for all the sources (black), for `1C' sources (red), for `2-3C' sources (green) and for sources with more than 3 components (blue, labelled `$>$3C'). The errors on median values are given by $1.253\times{\rm rms}/\sqrt{N}$, where rms is the standard deviation around the mean and $N$ is the number of detected sources \citep[cf.][]{Arkin1970}.}
\label{fig:PPolNCVsfreq}
\end{figure}

\subsection{Polarization angle: cm- and mm-wavelength regime behaviour}\label{sect:CP}
The polarization angle was calibrated setting the parameter `xycorr' in the MIRIAD task ATLOD which applies phase corrections provided by a noise diode mounted on one antenna feed. Partridge et al. (2016) found that the polarization angles measured by ATCA in this way agree with those measured by \textit{Planck} based on the CMB dipole measurements to within $\pm 2^\circ$.

\citet{Galluzzi2017} found evidence of non-zero Faraday rotation for only $2$ objects (over a total of $53$), since for the overwhelming majority of the sources the dependence of the rotation measure (RM) with $\lambda^2$ has a complex behaviour. Only $9$ objects of our larger sample can be described by a linear RM--$\lambda^2$ relation over the our full frequency range ($2.1$-$38\,$GHz). For these sources RM estimates are between $-72$ and $57\,$rad/m$^2$, with $4$ cases compatible with a low ($\sim \pm 10\,$rad/m$^2$) or a null rotation.

Exploiting our larger frequency range, we can identify two regimes for the RM vs $\lambda^2$ relation, one at cm-wavelengths and the other at mm-wavelengths. We have investigated this more complex scenario by fitting the polarization angle as a function of the $\lambda^2$ separately for the two regimes (from $2.1$ to $9\,$GHz and from $18$ to $38\,$GHz). We required at least three measured polarization angles in each regime to perform the fit via the IDL `linfit' procedure. A fit was regarded as acceptable when the reduced $\chi^2 < 2$ (probability $>0.1$). We obtained $\sim 40\%$ and $\sim 57\%$ successful fits for the low and high frequency regimes, respectively. The corresponding median values of the reduced $\chi^2$ are $0.37$ and $0.69$, respectively.

The medians and quartiles at cm- and mm-wavelengths are reported in Table~\ref{tab:RMLH} both for all objects for which acceptable fits were obtained and for the `1C', `2-3C' and `$>$3C' types, defined in sub-section~\ref{sect:LP}). We warn the reader that the error associated to the estimated RMs can be large especially at the higher frequencies because of its dependence on $1/\lambda^2$. Typical uncertainties are of about $9\%$ and $32\%$ at low and high frequencies, respectively; thus while at the lower frequencies only $2$ ($\sim 5\%$) of the estimated RMs are compatible with a null rotation at the $1\,\sigma$ significance level, this fraction raises to $15\%$ at the higher frequencies.

The median \textit{observed} (i.e. uncorrected for the effect of redshift) values of the RM in the low frequency regime are $\sim 40\,$rad/m$^2$ irrespective of the spectral type. At high frequencies they are larger for the whole sample (by a factor $\sim 15$, i.e. $\sim 600\,$rad/m$^2$) and for `1C' or `2-3C' objects, and much larger for the $>$3C objects ($\sim 1100\,$rad/m$^2$). Large values of RMs for multi-component sources were previously reported by \citet{Pasetto2016} who suggested that the youngest, highest frequency components can be characterised by a really dense and/or a magnetised medium that strongly rotates the polarisation angle.

So far we dealt with \textit{observed} RMs, ${\rm RM}_{\rm obs}$. The RM at the source, ${\rm RM}_{\rm AGN}$, are related to ${\rm RM}_{\rm obs}$ by \citet{JohnstonHollitt2004}:
\begin{equation}
{\rm RM}_{\rm obs}=\frac{{\rm RM}_{\rm AGN}}{(1+z)^2}+{\rm RM}_{\rm Gal}+{\rm RM}_{\rm ion},
\label{equ:TotRMcorrred}
\end{equation}
where RM$_{\rm Gal}$ and ${\rm RM}_{\rm ion}$ are the contributions of our own Galaxy  and of Earth's ionosphere, respectively.   
The Galactic contribution can typically vary from $-300$ to $300\,$rad/m$^2$ depending on the line of sight. Our sample is located in a region around the Southern Ecliptic Pole and we adopt the Galactic Faraday rotation map provided by \citet{Oppermann2014} to get the appropriate correction for each object of our sample.
The ionospheric contributions is found to be typically $\lesssim 5\,$rad/m$^2$ \citep{JohnstonHollitt2004} and can thus be safely neglected.

We have found redshifts in the AT20G catalogue (Mahony et al. 2011), and complemented them searching in the NED (NASA/IPAC Extragalactic Database) database. For sources with redshift ($44$) we have computed also ${\rm RM}_{\rm AGN}$ (see Table~\ref{tab:RMLH}). The median low-frequency value is around $90\,$rad/m$^2$. At high frequencies there seems to be a strong increase of the median ${\rm RM}_{\rm AGN}$ from `1C' to `2-3C' to `$>$3C' objects (median ${\rm RM}_{\rm AGN}$ of $\simeq 700$, $\simeq 2400$ and $\simeq 4000\,$rad/m$^2$), but the small numbers of `1C' and `$>$3C' objects prevents any firm conclusion. It is, however, remarkable that the large RMs of `$>$3C' objects echo the decrease of their median polarization fraction at mm wavelengths (sub-sect.~\ref{sect:LP}).

\begin{table*}
\caption{Median plus I and III quartile values of cm-wavelengths (upper table) and mm-wavelengths (lower table) RMs. In each table the upper set of values refers to the observed RMs while the lower set refers to the RMs at the source for the subset of sources for which redshift measurements are available. In parenthesis are the numbers of sources in each group. Whenever the number of objects is $< 10$ we provide only the median value. RMs are in rad/m$^2$.}
\label{tab:RMLH}
\begin{tabular}{|c|c|c|c|}
\hline
All sample ($42$)& 1C ($3$)&2-3C ($31$)& $>$3C ($8$)\\
\hline
\begin{tabular}{ccc}I&med&III\\18&37&58\end{tabular}&\begin{tabular}{ccc}I&med&III\\-&60&-\end{tabular}&\begin{tabular}{ccc}I&med&III\\15&34&53\end{tabular}&\begin{tabular}{ccc}I&med&III\\-&37&-\end{tabular}\\
\hline
All sample ($23$)& 1C ($2$)& 2-3C ($18$)& $>$3C ($3$)\\
\hline
\begin{tabular}{ccc}I&med&III\\40&94&244\end{tabular}&\begin{tabular}{ccc}I&med&III\\-&$335$&-\end{tabular}&\begin{tabular}{ccc}I&med&III\\46&84&220\end{tabular}&\begin{tabular}{ccc}I&med&III\\-&122&-\end{tabular}\\
\hline
\end{tabular}
\begin{tabular}{|c|c|c|c|}
\hline
All sample ($59$)& 1C ($4$)&2-3C ($50$)& $>$3C ($5$)\\
\hline
\begin{tabular}{ccc}I&med&III\\225&635&1397\end{tabular}&\begin{tabular}{ccc}I&med&III\\-&342&-\end{tabular}&\begin{tabular}{ccc}I&med&III\\283&637&1397\end{tabular}&\begin{tabular}{ccc}I&med&III\\-&1141&-\end{tabular}\\
\hline
All sample ($27$)& 1C ($2$)& 2-3C ($22$)& $>$3C ($3$)\\
\hline
\begin{tabular}{ccc}I&med&III\\679&2300&5252\end{tabular}&\begin{tabular}{ccc}I&med&III\\-&$742$&-\end{tabular}&\begin{tabular}{ccc}I&med&III\\716&2351&5191\end{tabular}&\begin{tabular}{ccc}I&med&III\\-&4022&-\end{tabular}\\
\hline
\end{tabular}
\end{table*}

\subsection{Circular polarization}
\label{subsec:CirPol}
The circularly polarized emission is weak, typically $\lesssim 0.1\%$ \citep{Rayner2000}, but potentially very interesting because its measurements may permit to gain information on various properties of jets, such as magnetic field strength and topology, the
net magnetic flux carried by jets (and hence generated in the central engine), the energy spectrum of radiating particles, and the jet composition, i.e. whether jets are mainly composed of electron-positron pairs or electron-proton plasma \citep{RuszkowskiBegelman2002}.

The most obvious candidate for explaining circular polarization of compact radio sources is intrinsic emission, but the expected level under realistic conditions appears to be too low to explain the observed polarization \citep{WardleHoman2003}. \citet{Pacholczyk1973} pointed out that magnetic fields computed from the circular polarization, assuming that it is intrinsic, are usually so high as to cause a turnover in the intensity spectrum through synchrotron self-absorption at a considerably higher frequency than is actually observed. The most promising mechanism is Faraday conversion, a birefringence effect that converts linear to circular polarization \citep{RuszkowskiBegelman2002, WardleHoman2003}.

At only two frequencies (5.5 and 9 GHz) more than 50\% of the sources were detected and thus median values of the circular polarization fractions, $m_{\rm V}$, could be determined. We find $m_{\rm V\,median}=(0.23\pm 0.01)\%$ and $(0.27\pm 0.02)\%$, respectively. For comparison, the median $m_{\rm V}$ for the \citet{Rayner2000} sample, selected at 4.85\,GHz, estimated from the data in their Table~3, is $\simeq (0.05\pm 0.02)\%$. Our larger median values may be due to the fact that, because of the higher selection frequency, the overwhelming majority of objects in our sample are blazars; \citet{Rayner2000} have found that these objects have larger circular polarization fractions than radio galaxies that comprise a significant fraction ($\simeq 25\%$) of their sample.

\section{Source counts in polarization}
\label{sec:SouCou}

\begin{figure}
\includegraphics[width=\columnwidth]{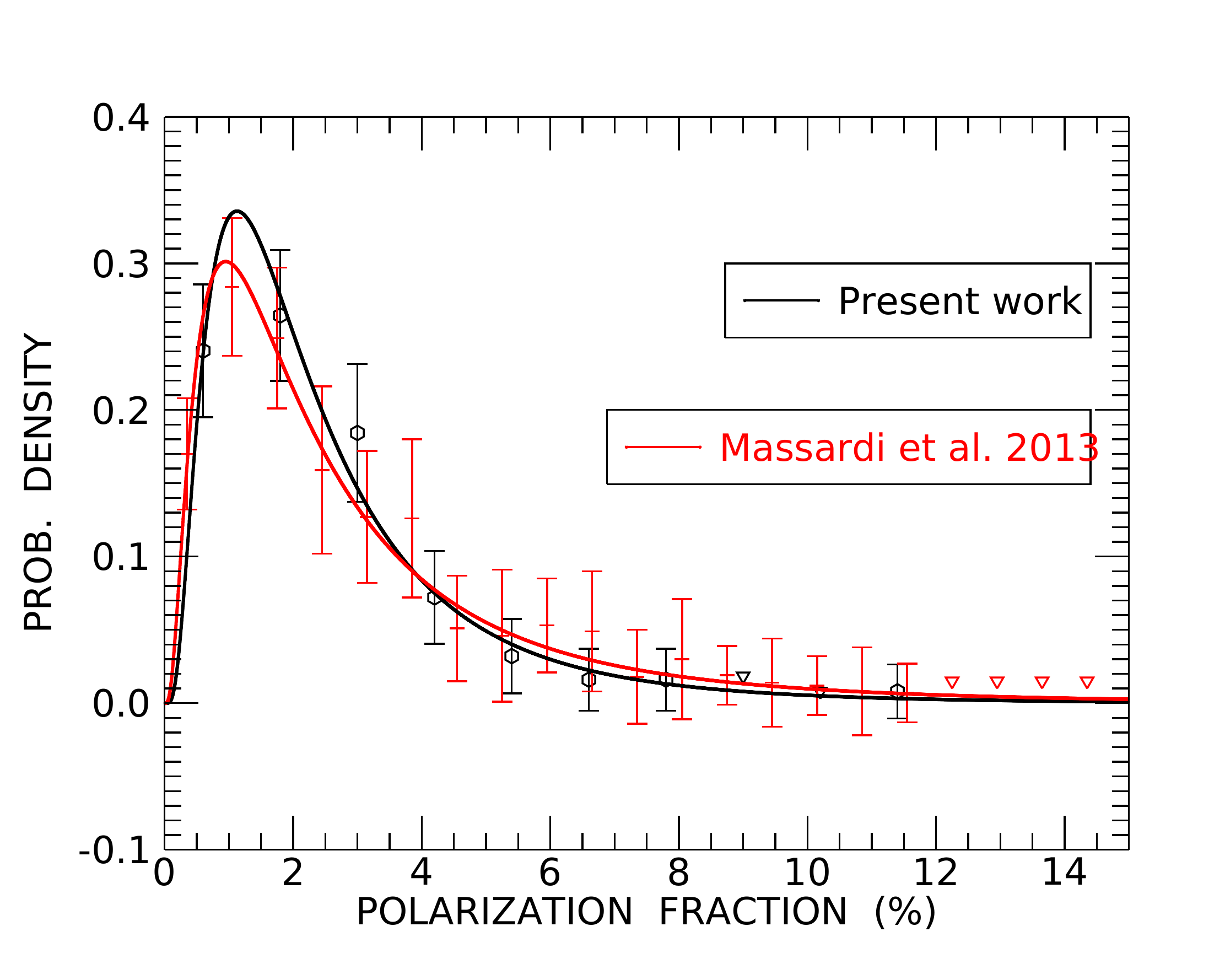}
\caption{Distribution of the polarization fraction at $18\,$GHz. Errors and upper limits correspond to a $1\sigma$ level. The black circles refer to the sample studied in this paper, the red pluses to the full AT20G bright sample studied in Massardi et al. 2013. The corresponding fit by a log-normal distribution for each dataset is reported with a solid lines of the same color.}
\label{fig:polfrac18G}
\end{figure}

\begin{figure}
\includegraphics[width=\columnwidth]{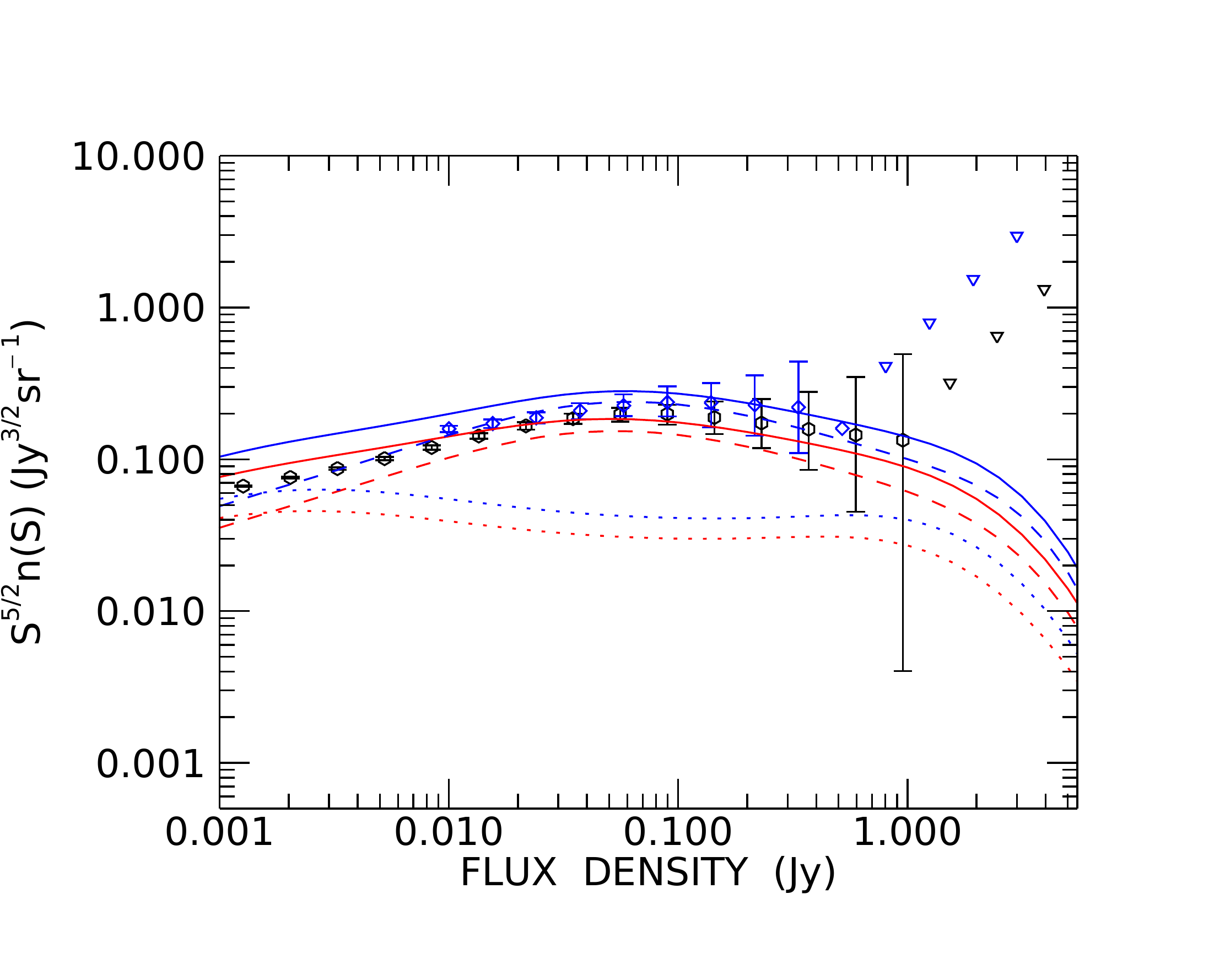}
\caption{Differential source counts at $20\,$GHz in polarization obtained in this paper plotted with black circles (black downward triangles are for upper limits). Also shown, for comparison, are the estimates by \citet{Massardi2013} using the polarimetric data from their own survey, somewhat shallower than the present one ($S_{20\,{\rm GHz}}>500\,$mJy) combined with the full AT20G catalogue (blue diamonds and blue downward triangles for upper limits). The curves show the predictions of the \citet{TucciToffolatti2012} model: blue curves for the `conservative' case and red curves for the `optimistic' case. The solid lines represent the total number counts; the dotted lines are for steep-spectrum sources (classified at low frequencies); the dashed lines are for flat objects (flat-spectrum radio quasars, i.e. FSRQs and BL Lacs).}
\label{fig:DifSouCou18G}
\end{figure}

Figure~\ref{fig:DifSouCou18G} shows the source counts in polarization at $20\,$GHz obtained through the convolution of the total intensity differential source counts reported by the model De Zotti et al. (2005) with our distribution of polarization fractions at $18\,$GHz. In the Table~\ref{tab:polfrac18G} and in the Figure~\ref{fig:polfrac18G} we report the observed distribution (black circles): in each bin uncertainties are derived assuming a Poisson statistics, following the indications of Gehrels (1986). The solid line is the fit assuming a log-normal distribution
\begin{equation}
f(\Pi)={\rm const}\cdot \frac{1}{\sqrt{2\pi}\sigma \Pi}\exp^{-\frac{1}{2}ln\left(\Pi / \Pi_m\right)^2/\sigma^2},
\label{equ:lognordis}
\end{equation}
where const$=0.96$, $\sigma=0.76$ and $\Pi_m=2.00$, i.e. the median value of the distribution. The reduced $\chi^2$ value is $0.21$. In the Table~\ref{tab:DifSouCou18G} and in the figure~\ref{fig:DifSouCou18G} (black circles) we plot the differential source counts in polarization, following the recipe reported by Tucci \& Toffolatti (2012): since there is no evidence of a correlation between the total intensity flux density and the polarization fraction, the number counts $n(P)\equiv dN/dP$ can be determined by
\begin{equation}
n(P)=\int_{S_0=P}^\infty {\mathcal{P}\left(m=\frac{P}{S}\right)n(S)\frac{dS}{S}},
\label{equ:difsoucouP}
\end{equation}
where $n(S)$ is the assumed source counts in total intensity, $\mathcal{P}$ is the probability density distribution for the polarization fraction $m$, i.e. $\Pi/100$. Note that in each bin in $P$ the integration over $S$ is truncated at $S_0=P$, which corresponds to the maximum degree of the polarization fraction (i.e. $m=1.0$). We compare our results with source counts provided by Massardi et al. (2013, blue diamonds) via a MCMC simulation of the whole AT20G catalogue (Massardi et al. 2011a), as well as with the Tucci \& Toffolatti model (2012, red and blue lines, which refer to the lower and upper level expected, respectively). Since our sample is mainly composed by blazars (BL Lacs and FSRQs), which typically are labelled as `flat' and represent the dominant population at $20\,$GHz (dashed lines), we expect and find a good agreement with the limits on the total source counts provided by the model. Hence, given the assumptions of \citet{TucciToffolatti2012} on the median polarization fraction of steep--spectrum radio sources (presented in their Table 4), that are higher than our current findings (see our Figure 5, for a comparison), their overestimation of source number counts in polarization below $10\,$mJy can be (at least partially) explained. Note that eq.~(\ref{equ:difsoucouP}) assumes independence of the polarization fraction from the total flux density. However this assumption can be broken as another population, namely steep-spectrum sources, with different polarization properties, becomes increasingly important with decreasing flux density.

\begin{table}
\caption{Distribution of the polarization fractions at $18\,$GHz for the full sample.}
\label{tab:polfrac18G}
\begin{tabular}{cccc}
\hline
$\Pi$ (per cent)& Probability & lower & upper\\
 & & uncert. & uncert.\\
\hline
0.600 & 0.2404 & 0.0453 & 0.0453\\
1.800 & 0.2644 & 0.0446 & 0.0446\\
3.000 & 0.1843 & 0.0381 & 0.0470\\
4.200 & 0.0721 & 0.0222 & 0.0317\\
5.400 & 0.0321 & 0.0153 & 0.0253\\
6.600 & 0.0160 & 0.0104 & 0.0211\\
7.800 & 0.0160 & 0.0104 & 0.0211\\
9.000 & $<$0.01843 & & \\
10.200 & $<$0.00801 & & \\
11.400 & 0.0080 & 0.0066 & 0.0184\\
\hline
\end{tabular}
\end{table}

\begin{table}
\caption{Euclidean normalized differential source counts at $20\,$GHz in polarization obtained in this paper via the convolution of the distribution of the polarization fraction at $18\,$GHz with the De Zotti model (2005).}
\label{tab:DifSouCou18G}
\begin{tabular}{cccc}
\hline
$\log{[P({Jy})]}$ & $S^{5/2}n(S)$  (Jy$^{3/2}$sr$^{-1}$)& lower & upper \\
 & & uncert. & uncert.\\
\hline
-2.897 & 0.0667 & 0.0007 & 0.0007\\
-2.692 & 0.0760 & 0.0011 & 0.0011\\
-2.486 & 0.0869 & 0.0017 & 0.0017\\
-2.281 & 0.1011 & 0.0025 & 0.0025\\
-2.075 & 0.1198 & 0.0039 & 0.0039\\
-1.870 & 0.1426 & 0.0061 & 0.0061\\
-1.664 & 0.1662 & 0.0094 & 0.0094\\
-1.459 & 0.1856 & 0.0142 & 0.0142\\
-1.253 & 0.1978 & 0.0209 & 0.0209\\
-1.048 & 0.1987 & 0.0299 & 0.0299\\
-0.842 & 0.1886 & 0.0417 & 0.0519\\
-0.637 & 0.1734 & 0.0549 & 0.0766\\
-0.431 & 0.1580 & 0.0726 & 0.1199\\
-0.226 & 0.1447 & 0.0996 & 0.2034\\
-0.020 & 0.1337 & 0.1297 & 0.3606\\
0.185 & $<$0.31886 & & \\
0.391 & $<0$.64841 & & \\
0.596 & $<$1.31855 & & \\
\hline
\end{tabular}
\end{table}

\section{Conclusions}
\label{sec:discusseconcl}

We have presented and discussed high sensitivity polarimetric observations in $7$ bands, centered at $2.1$, $5.5$, $9$, $18$, $24$, $33$ and $38\,$GHz, of a complete sample of 104 extragalactic sources with $S_{20\rm GHz}\ge 200\,$mJy in the AT20G catalogue. The r.m.s error in the polarized flux density is $0.6\,$mJy at $\nu \ge 5.5\,$GHz and $1\,$mJy at $2.1\,$GHz, due to the heavy RFI contamination.

Polarization measurements in the range $5.5-38\,$ GHz for $53$ objects of the sample were reported by \citet{Galluzzi2017}. The measurements for the other $51$ sources are new, as are the $2.1\,$GHz measurements for the full sample of $104$ sources. The $53$ sources were re-observed at $5.5$ and $9\,$GHz, while we managed to repeat the observations at $18$, $24$, $33$ and $38\,$GHz only for $20\%$ of them. The previous measurements at $33$ and $38\,$GHz were re-calibrated using the updated model for the flux density absolute calibrator, PKS1934-638, that was not available for the earlier analysis.

The observational determination of the continuum spectra has been extended by exploiting the GLEAM survey data at $20$ frequencies between $72$ and $231\,$MHz \citep{HurleyWalker2017}, available for $89$ ($\simeq 86\%$) of our sources. For these sources we have the unparalleled coverage of $2.7$ decades in frequency.

The total intensity data from $5.5$ to $38\,$GHz could be interpreted in terms of a single emission region \citep{Galluzzi2017}. A joint analysis of the more extended total intensity spectra (from $72\,$MHz to $38\,$GHz) presented here and of the polarization spectra reveals a more complex astrophysics. About $93\%$ of our sources show clear indications of at least two emission components, one (or sometimes more) dominating at the higher frequencies and self-absorbed at a few GHz, and another one, generally steeper, emerging at lower frequencies. The most straightforward interpretation of these results is in terms of recurrent activity, with the extension (and, correspondingly, the age) of emission components increasing with decreasing frequency, i.e. with younger components showing up at higher frequencies.

There is no evidence of trends of the linear polarization fraction with the frequency for the full sample and for the single component (`1C') sub-set. However, sources with 2 or 3 components (`2-3C') have a minimum of the polarization fraction at $\simeq 9\,$GHz, consistent with different emission components at lower and higher frequencies. For for multi-component (`$>$3C')  objects there is a hint of a \textit{decrease} with increasing frequency of the polarization fraction, although the statistics is very poor. 

Further indications of different origins for the low- and high-frequency emissions come from our analysis of rotation measures. The data suggest two regimes for the RM vs $\lambda^2$ relation, one at cm-wavelengths, with typical \textit{intrinsic} RM of 
$\sim 90\,$rad/m$^2$, and the other for mm-wavelengths with median intrinsic $\hbox{RM}\sim  2000\,$rad/m$^2$ (but with very large errors). The `$>$3C' seem to have very high RMs ($\sim 4000\,$rad/m$^2$). Again the statistics is very poor but it is suggestive that, at mm wavelengths, the large RMs echo the low polarization fraction.

Our high sensitivity polarimetry has allowed a $5\,\sigma$ detection of the weak circular polarization for $\sim 38\%$ of data. The measured values of Stokes' $V$, while much lower than the linear polarization amplitude, are much higher than expected for the intrinsic circular polarization of synchrotron emission corresponding to the typical magnetic field intensities in radio emitting regions. This is consistent with previous conclusions in the literature that circular polarization is predominantly produced by Faraday conversion of linear polarization.

Finally we have presented a new estimate of the counts in linear polarization at $18\,$GHz derived from the convolution of the distribution of polarization fractions for our sample with the model for total intensity source counts by De Zotti et al. (2005) that, thanks to the high sensitivity of our data, allows to reach deeper polarized flux density levels than obtained so far.

\section*{Acknowledgments}
We thank the anonymous referee for useful comments.
We acknowledge financial support by the Italian {\it Ministero dell'Istruzione, Universit\`a e Ricerca} through the grant {\it Progetti Premiali 2012-iALMA} (CUP C52I13000140001).
Partial support by ASI/INAF Agreement 2014-024-R.1 for the {\it Planck} LFI Activity of Phase E2 and by ASI through the contract I-022-11-0 LSPE is acknowledged.
We thank the staff at the Australia Telescope Compact Array site, Narrabri (NSW), for the valuable support they provide in running the telescope and in data reduction. The Australia Telescope Compact Array is part of the Australia Telescope which is funded by the Commonwealth of Australia for operation as a National Facility managed by CSIRO. AB acknowledges support from the European Research Council under the EC FP7 grant number 280127. VG thanks Rocco Lico for the useful discussions. VC acknowledges DustPedia, a collaborative focused research project supported by the European Union under the Seventh Framework Programme
(2007-2013) call (proposal no. 606824). The participating institutions are: Cardiff University, UK; National Observatory of Athens, Greece; Ghent University, Belgium; Université Paris Sud, France; National Institute for Astrophysics, Italy and CEA (Paris), France. LT and LB acknowledges partial financial support from the Spanish Ministry of Economy and Competitiveness (MINECO), under project AYA-2015-65887-P.

\bsp


\end{document}